\documentclass[acmsmall]{acmart}

\usepackage{graphicx}
\usepackage{tikz}
  
\usepackage{forest}
\usetikzlibrary{trees,positioning,shapes,shadows,arrows.meta}

\usepackage{amsmath,amsfonts}
\usepackage{algorithmic}
\usepackage{textcomp}
\usepackage{multirow}

\usepackage{tabulary}
\usepackage[para]{threeparttable}
\usepackage{array,booktabs,longtable,tabularx}
\newcolumntype{L}{>{\raggedright\arraybackslash}X}
\usepackage{ltablex}
\usepackage{siunitx}
\usepackage{float} 
\usepackage{url}
\usepackage{courier}
\usepackage{amsthm}
 
\usepackage{makecell}

\theoremstyle{definition} 

\newtheorem*{controlflowgraph*}{Control Flow Graph (CFG)}
\newtheorem*{fcg*}{Function Call Graph (FCG)}
\newtheorem*{dependence*}{Program Dependence Graph (PDG)}
\newtheorem*{syscall*}{System Call Graph}
\newtheorem*{entities*}{System Entity Graph}
\newtheorem*{flows*}{Network Flow Graph}

\newtheorem*{graph*}{Graph}
\newtheorem*{attributed*}{Attributed Graph}
\newtheorem*{hetero*}{Heterogeneous Graph}

\newtheorem*{cicandmal*}{CICAndMal2017 \cite{lashkari2018toward}}
\newtheorem*{cicmaldroid*}{CICMalDroid \cite{mahdavifar2020dynamic}}
\newtheorem*{androzoo*}{AndroZoo \cite{allix2016androzoo}}
\newtheorem*{drebin*}{Drebin \cite{arp2014drebin}}
\newtheorem*{malnet*}{MalNet \cite{freitas2020large}}

\newtheorem*{microsoft*}{Microsoft Malware Classification Challenge (MMCC) \cite{ronen2018microsoft}}
\newtheorem*{virushsharevirushtotal*}{VirusShare \cite{virusshare}, VirusTotal \cite{virustotal}}

\newtheorem*{phishtank*}{PhishTank \cite{phishtank}, OpenPhish \cite{openphish}}
\newtheorem*{alexa*}{TrancoTop1M\cite{pochat2018tranco}, AlexaTop1M \cite{alexa}}
\newtheorem*{xssed*}{XSSed\cite{xssed}}
\newtheorem*{petrak*}{Petrak\cite{petrak}, GeeksOnSecurity\cite{geeksonsecurity}}
\newtheorem*{caida*}{CAIDA \cite{caida}}
\newtheorem*{ctu*}{CTU-13 \cite{garcia2014empirical}}
\newtheorem*{cic2017*}{CIC-IDS2017 \cite{sharafaldin2018toward}}
\newtheorem*{cicdos*}{CIC-DoS-2017 \cite{jazi2017detecting}}

\newcommand\addtag{\refstepcounter{equation}\tag{\theequation}}

\def\BibTeX{{\rm B\kern-.05em{\sc i\kern-.025em b}\kern-.08em
    T\kern-.1667em\lower.7ex\hbox{E}\kern-.125emX}}
\AtBeginDocument{%
  \providecommand\BibTeX{{%
    \normalfont B\kern-0.5em{\scshape i\kern-0.25em b}\kern-0.8em\TeX}}}

\setcopyright{acmcopyright}
\copyrightyear{2023}
\acmYear{2023}
\acmDOI{XXXXXXX.XXXXXXX}

\acmJournal{JACM}
\acmVolume{1}
\acmNumber{1}
\acmArticle{1}
\acmMonth{3}

\begin{document}

\title{A Survey on Malware Detection with Graph Representation Learning}


\author{Tristan Bilot}
\affiliation{%
  \institution{Iriguard}
  \streetaddress{5 Rue Bellini}
  \city{Puteaux}
  \country{France}
}
\affiliation{%
    \institution{Université Paris-Saclay, CNRS, Laboratoire Interdisciplinaire des Sciences du Numérique}
  \city{Gif-sur-Yvette}
  \country{France}
}
\affiliation{%
    \institution{LISITE Laboratory, ISEP}
  \streetaddress{10 Rue de Vanves}
  \city{Issy-les-Moulineaux}
  \country{France}
}
\email{tristan.bilot@universite-paris-saclay.fr}

\author{Nour El Madhoun}
\affiliation{%
  \institution{LISITE Laboratory, ISEP}
  \streetaddress{10 Rue de Vanves}
  \city{Issy-les-Moulineaux}
  \country{France}
}
\affiliation{%
  \institution{Sorbonne Université, CNRS, LIP6}
  \streetaddress{4 place Jussieu}
  \city{Paris}
  \country{France}
}
\email{nour.el-madhoun@isep.fr}

\author{Khaldoun Al Agha}
\affiliation{%
 \institution{Université Paris-Saclay, CNRS, Laboratoire Interdisciplinaire des Sciences du Numérique}
 \city{Gif-sur-Yvette}
 \country{France}
 }
 \email{alagha@lisn.fr}

\author{Anis Zouaoui}
\affiliation{%
  \institution{Iriguard}
  \streetaddress{5 Rue Bellini}
  \city{Puteaux}
  \country{France}
}
\email{a.zouaoui@iriguard.com}

\renewcommand{\shortauthors}{T. Bilot et al.}

\begin{abstract}
Malware detection has become a major concern due to the increasing number and complexity of malware. Traditional detection methods based on signatures and heuristics are used for malware detection, but unfortunately, they suffer from poor generalization to unknown attacks and can be easily circumvented using obfuscation techniques. In recent years, Machine Learning (ML) and notably Deep Learning (DL) achieved impressive results in malware detection by learning useful representations from data and have become a solution preferred over traditional methods. More recently, the application of such techniques on graph-structured data has achieved state-of-the-art performance in various domains and demonstrates promising results in learning more robust representations from malware. Yet, no literature review focusing on graph-based deep learning for malware detection exists. In this survey, we provide an in-depth literature review to summarize and unify existing works under the common approaches and architectures. We notably demonstrate that Graph Neural Networks (GNNs) reach competitive results in learning robust embeddings from malware represented as expressive graph structures, leading to an efficient detection by downstream classifiers. This paper also reviews adversarial attacks that are utilized to fool graph-based detection methods. Challenges and future research directions are discussed at the end of the paper.
\end{abstract}



\maketitle

\section{Introduction}
\label{sec:introduction}

Malware, short for malicious software, is a generic term for unwanted programs designed to harm or exploit computer systems  \cite{bayer2006dynamic}. The detection of widespread malware such as ransomware, worms, Trojan horses or spyware, has become a major concern since their increase in both number and complexity \cite{ye2017survey}. Indeed, malware programs can appear in different forms and may be hidden under other trusted programs available on the most used platforms such as Android, Windows or even the Web. Unaware users are frequently fooled by authors of malware and important efforts have been spent to prevent these threats. Traditional detection techniques mainly rely on signatures and heuristics, where malware is detected by comparing it to existing malware or known malicious patterns. However, those methods are known to suffer from poor generalization to unknown attacks or variants and can be easily circumvented using obfuscation techniques \cite{aslan2020comprehensive}. Other behavior-based methods tend to perform better by further analyzing the malware and evaluating its intended actions before executing it. However, such techniques appear to be very time-consuming \cite{kuchler2021does}. Over the last decade, Machine Learning (ML) and notably Deep Learning (DL) have sparked a sea change in a variety of fields, including cybersecurity, by allowing the model to learn from data and adapt to new patterns. This ability to adapt makes these methods well-suited to a number of tasks, including malware detection, as shown by the growing number of papers that apply ML to this problem \cite{liu2020review}.

Despite the progress made with these learning-based methods, malware detection remains a challenging task, as malware authors continue to make their techniques evolve, with the aim to evade detection. In an attempt to outperform current ML and DL methods that learn from traditional Euclidean data, graph representation learning has emerged as a promising alternative to capture complex patterns in malware programs represented as graphs. Indeed, a growing number of fields are benefiting from these graph-based learning methods and obtaining state-of-the-art results \cite{zhou2020graph}, as graph structures offer even more semantic information by encoding spatial relations and connectivity between entities.

Current studies on malware detection using machine learning are mainly based on the review of traditional ML and DL techniques applied to structured data. However, more and more recent papers tend to use graph representation learning in their approaches, and to the best of our knowledge, there is no literature review that specifically focuses on these techniques applied to malware detection.

This survey is a first attempt to shape the research area of malware detection with graph representation learning, by providing a comprehensive review of current approaches. Specifically, we present in this paper the following contributions:

\begin{itemize}
    \item An overview of common representations that are used to model malware as graphs as well as techniques to extract these graph structures from raw malware data.
    
    \item A comprehensive summary of the state of the art papers, grouped according to the most common types of graphs, namely: Control Flow Graph (CFG), Function Call Graph (FCG), Program Dependence Graph (PDG), system call graph, system entity graph and network flow graph. We also propose a general architecture under which a majority of works can be abstractly summarized.
    
    \item A review of the adversarial attacks that are used against GNN-based malware detection techniques, along with a discussion on the challenges that may be encountered as well as future research directions and conclusions. In particular, we show that the works presented in this paper are very recent and that many promising directions remain unexplored.
\end{itemize}


The paper is organized as follows. In section \ref{sec:related}, we introduce related works and further explain the contributions of our paper. In section \ref{graph-ml-section}, we provide background knowledge on graphs and present the fundamentals of graph representation learning and Graph Neural Networks (GNNs). Section \ref{sec:malware_detection} discusses the techniques used to extract graph-structured data from malware as well as the general architecture used for their detection with representation learning techniques. Sections \ref{sec:android} to \ref{sec:web} review the state-of-the-art papers for the detection of Android, Windows and Web malware, respectively. Section \ref{adversarial} discusses the robustness of GNN-based detection systems against adversarial attacks. In section \ref{sec:future}, some challenges and an overview of future research are discussed. The last section \ref{conclusion} concludes this paper.

\section{Related Works} \label{sec:related}

In existing literature, several studies have been published that aim to review malware detection using standard ML and DL techniques. The authors of the paper \cite{aslan2020comprehensive} have conducted a comprehensive review on malware detection. They first present the problem of malware detection, as well as the various challenges that can be encountered and the techniques used to overcome them. They also review a significant number of papers based on traditional methods such as signatures, behaviors and heuristics, but also cover some ML-based methods.

The paper \cite{naway2018review} proposes to review the deep learning models employed in Android malware detection, focusing on the analysis of the strengths and weaknesses of these models. The literature is comprehensively summarized by providing useful information about each research work, including the analysis method, features, models used and their performance, and input datasets. The proposed survey in the article \cite{qiu2020survey} covers a wide variety of deep neural models used for Android malware detection and mentions few graph-based methods using control flow graphs \cite{yang2014droidminer,atici2016android} and App-API graphs \cite{hou2018make,hou2017hindroid}.

Authors in \cite{liu2020review} surveyed the traditional ML techniques employed in Android malware detection and explain the commonly employed ML tasks such as data acquisition, data preprocessing, and feature selection.  In the paper \cite{wang2020review}, a large category of deep learning methods using static, dynamic and hybrid analysis is reviewed. Important information is provided regarding the input features that can be extracted from APKs, as well as the most commonly used datasets for both benignware and malicious Android software. 

The survey \cite{singh2021survey} analyzes traditional ML methods in a general approach for malware detection based on executable files. Representation learning methods applied to cybersecurity are reviewed in the study \cite{usman2019survey}, with few mentions to malware detection. More recently, the paper \cite{gopinath2023comprehensive} also reviewed DL methods applied to the detection of mobile malware, Windows malware, IoT malware, Advanced Persistent Threats (APTs) and Ransomware.

Regarding graph representation learning and GNN-based methods, the work \cite{warmsley2022survey} surveys GNN techniques employed for malware analysis with a focus on the prediction explainability. Other surveys review the applications of GNNs \cite{zhou2020graph,zhang2020deep,wu2020comprehensive,chen2020graph} but none of them mention malware detection.

Indeed, after extensive research and to the best of our knowledge, the literature on malware detection using ML and DL techniques is widely covered and documented but it is still missing a review dedicated to graph ML and graph representation learning methods. Our paper focuses on analyzing recent research studies based on such methods for malware detection, starting from the extraction of graph-structured data using reverse engineering tools, to the classification of malware based on graph embeddings. Our goal is to provide the necessary knowledge to researchers interested in the application of ML to graph-structured malware, and to contribute to the advancement of this field.

\section{Background} 
\label{graph-ml-section}
In this section, we introduce the fundamentals about graphs along with the graph representation learning techniques leveraged to learn from these structures. We first discuss the properties of graphs and then explain differences between traditional Deep Learning and graph representation learning, along with the types of GNNs that are frequently employed.

\subsection{Graph Structures} \label{sec:graphs}

Graphs are useful data structures to model the interactions between the entities of a complex system. They possess a great expressiveness and can represent any connected systems using only two abstract objects, which are nodes and edges. 

\begin{graph*}
A graph can be denoted as $G=(V, E)$ where $V={v_1, ..., v_N}$ is a set of $N=|V|$ nodes (i.e. entities) and $E={e_1, ..., e_M}$ is a set of $M=|E|$ edges, namely the relations between entities. Edges in the graph can either be directed (e.g. a process $a$ forks another process $b$), or undirected (e.g. a bi-directional communication flow between two clients). By default, such graphs only represent a topology by incorporating the relations between different objects and do not store any local information.
\end{graph*}

\begin{attributed*}
Attributed graphs attach additional features to the elements of the graph, leading to a more detailed representation. A node-attributed graph assumes function $F_n: V \longrightarrow \mathbb{R}^{d_n}$ to map each node to a feature vector of $d_n$ elements. Similarly, an edge-attributed graph assumes function $F_e: E \longrightarrow \mathbb{R}^{d_e}$ to map every edge to a vector of $d_e$ features. Node and edge features can be conveniently described in a matrix format, where $X$ usually represents the node feature matrix and $X_e$ is the edge feature matrix. Furthermore, the structure of the graph is mostly designated by an adjacency matrix $A$.
\end{attributed*}

\begin{hetero*}
In many cases, the relations between graph objects become more complex, involving multiple types of modalities. These representations can be modeled with heterogeneous graphs, by introducing two mapping functions $\phi_v:V \rightarrow T_v$ and $\phi_e:E \rightarrow T_e$ that respectively map to a node type in $T_v$ and an edge type in $T_e$.
\end{hetero*}

Although other graph structures exist, current state-of-the-art graph-based malware detection methods are mostly based on these representations.

\subsection{Learning on Graphs}
Since nodes in a graph are inherently connected, they are not considered independent and uniformly distributed. For these reasons, traditional ML models cannot be directly applied on graphs, which suggests that specific techniques are required to deal with these interconnected structures. 

\subsubsection{Representation Learning}
A malware detection model must first go through a training procedure where it learns parameters based on a large number of training samples, in order to approximate a relationship function between the input feature space and the output binary label. Representation learning aims at learning an intermediate function $f$ formulated as $f: X \rightarrow \mathbb{R}^d$, which maps the input feature space $X$ to an embedding space $\mathbb{R}^d$ that retains essential information from raw input features. Embedding representations can then be leveraged in downstream tasks such as learning word relationships \cite{mikolov2013distributed}, learning the representation of objects in images \cite{vincent2010stacked} or learning translation of language \cite{wu2016google}. In malware detection, representation learning aims at creating embeddings from input data such as program code. The embeddings are then converted into a distribution that either indicates a probability to be a malware (binary classification) or to belong to a determined malware category or malware family  (multi-class classification).

\subsubsection{Graph Representation Learning}
Standard representation learning techniques are not suited to deal with data generated from non-Euclidean domain space such as graphs. For instance, regular Convolutional Neural Networks (CNNs) \cite{lecun1995convolutional} and Recurrent Neural Networks (RNNs) \cite{hochreiter1997long} are unable to perform traditional convolutions or recurrent operations on graphs as the notion of Euclidean distance cannot be applied. Graph representation learning \cite{hamilton2017representation}, on the other hand, is a specific area of ML that aims to learn embedding representations from graph-structured data. This involves learning embeddings from nodes, edges, or graphs in a way that ensures that objects with similarities in feature space have similar representations in embedding space. The proximity between learned representations can then be leveraged in different downstream tasks. In the field of cybersecurity, tasks such as node classification, edge classification and graph classification are frequently used.
Node classification aims at finding a label for a specific object in the graph such as detecting a botnet node in a network \cite{zhou2020automating, carpenter2021detecting, zhao2020multi}, whereas edge classification is applied to assign a label to a relation or event, such as detecting a malicious authentication request \cite{bowman2020detecting,kingeuler}. On the other hand, graph classification maps the whole graph to a label. This task is largely used in malware detection in cases where the goal is to predict the label of a binary represented as a graph \cite{norouzian2021hybroid, xu2021hybrid, fairbanks2021identifying, zhu2018android, feng2020android, cai2021learning, xu2021detecting, vinayaka2021android, errica2021robust, catal2021malware, wu2022deepcatra, lo2022graph, gunduz2022malware, lu2022robust, yumlembam2022iot, xu2021android, john2020graph, busch2021nf, liu2023nt, yan2019classifying, xu2021hawkeye, wang2021hierarchical, ling2022malgraph, jiang2018dlgraph, oliveira2019behavioral, zhang2020spectral, li2022intelligent, li2022dmalnet, hung2019malware, wang2019heterogeneous, liu2022fewm}. It is also possible to work at the sub-graph level to detect areas in the graph that are responsible for the prediction done by a predictive model \cite{herath2022cfgexplainer}.

In literature, the first methods for graph representation learning based on graph embedding  are mostly relying on random walks, where the co-occurrence of nodes is preserved. DeepWalk \cite{perozzi2014deepwalk} was the first method to leverage the Skip-gram model \cite{mikolov2013distributed} to compute embeddings from nodes that co-occur in random walks. It learns node embeddings by optimizing a neighborhood preserving objective, using random walks and word embedding techniques. First, $n$ random walks are generated by randomly traversing the graph $n$ times. Each walk is composed of $k$ nodes, where $k$ is a hyperparameter representing the length of a random walk. Then each node tries to reconstruct neighboring nodes from its random walk using the Skip-gram model.

To fully learn the embeddings, node2vec \cite{grover2016node2vec} integrates a second-order biased random walk that captures local and global structures using Breadth First Search (BFS) and Depth First Search (DFS) algorithms.
Other methods such as LINE \cite{tang2015line} have also achieved great performance in learning embeddings from graphs. However, most of these techniques do not share parameters between nodes \cite{hamilton2017representation}, meaning that the model size grows linearly with the size of the graph. Moreover, these methods are highly dependent on the values of hyperparameters and tend to favor proximity information over structural information \cite{velickovic2019deep}. Another disadvantage of using these walk-based techniques is that they are generally transductive, meaning that a single graph is taken as input and that no inference is possible on unseen nodes or edges. Contrarily, inductive models take as input multiple graphs and can generalize to unseen examples.

\subsubsection{Graph Neural Networks} \label{section:gnn}
Recent graph representation learning approaches tend to be inspired from the Graph Neural Network (GNN) model \cite{gori2005new}\cite{scarselli2008graph}, which is the origin of the first application of deep neural networks to graph-structured data. Although deep learning on graphs has been democratized fairly recently, the first GNN \cite{gori2005new} dates back to 2005 and is originally inspired from RNNs. In recent years, the popularity of deep learning has led to the emergence of new methods involving spectral and spatial convolution methods applied to graph structures, making it possible to take advantage of both the expressive structure of graphs and the power of representation learning. Spectral GNNs such as ChebNet \cite{defferrard2016convolutional} exploit the Laplacian matrix eigen decomposition in Fourier space to analyze the underlying structure of the graph. On the other hand, spatial GNNs such as Graph Convolutional Network (GCN) \cite{kipf2016semi}, GraphSAGE \cite{hamilton2017inductive}, Deep Graph Convolutional Neural Network (DGCNN) \cite{zhang2018end}, and Graph Attention Network (GAT) \cite{velivckovic2017graph} work directly on the adjacency matrix and capture the local neighborhood of the nodes in the graph domain, which avoids the time-consuming switch in spectral domain. GCN captures both feature and local substructure information by propagating the information along the neighboring nodes within the graph. DGCNN also leverages convolutions but is specifically designed for the graph classification task. GraphSAGE provides an inductive solution that can scale to large graphs by sampling the neighbors during message-passing. Finally, GAT leverages the attention mechanism \cite{bahdanau2014neural} to learn an importance weight for each neighboring node.

 Variants of these models have achieved state-of-the-art results in a variety of domains such as recommender systems \cite{wu2022graph}, traffic forecasting \cite{jiang2022graph} and drug discovery
 \cite{gaudelet2021utilizing}. However, GNN-based methods remain little used in cybersecurity compared to other domains where research is largely oriented in this direction.

\section{Malware Detection with Graphs}  \label{sec:malware_detection}
In the field of ML, malware detection usually consists in extracting features from an input binary file, which are leveraged by a downstream algorithm for classification. Malware detection with graph ML follows the same idea, with the only difference that a supplementary step is introduced after the feature extraction. This step consists in transforming the input features into a graph structure that will then be fed to the classifier. In this section, we describe ways to represent malware as expressive graph structures along with a general methodology to leverage graph representation learning in downstream malware detection tasks.

\subsection{Modeling Malware as Graphs}
In real-world scenarios, malware programs are usually compiled binaries that may be obfuscated to hide their malicious payload. Furthermore, a same malware could be written in multiple languages or using different hardware platforms. Therefore, we think that an optimal representation of a binary program should fulfill these conditions:
\begin{itemize}
    \item \textbf{Preserve the semantic of the program}: the actions resulting from the execution of the app should be captured by the data representation, in order to understand benign and malicious behaviors.
    \item \textbf{Be robust to obfuscation techniques}: the representation should capture the fundamental semantic of the program even if its code is obfuscated.
    \item \textbf{Be language- and platform-agnostic}: the representation should be  abstract enough to transcend the programming language and the platform on which the program is written.
\end{itemize}
Building a data representation that respects all previous conditions remains a challenging task due to the constant evolution of techniques employed by attackers. In the next section, we describe the analysis methods and graph structures commonly used for the representation of malware.

\subsubsection{Analysis Methods for Feature Extraction} \label{section:analysis}
Multiple analysis techniques are commonly employed to extract meaningful features from computer programs in an attempt to obtain an efficient representation \cite{ye2017survey}. Static analysis aims to analyze software without executing it, whereas dynamic analysis actually executes it to capture different levels of information. Both approaches have their own strengths and weaknesses. Static analysis is relatively cheap to perform and provides a comprehensive view of the program by considering all branches present in the code. However, it may not detect issues that only occur at runtime, such as memory leaks and race conditions. On the other hand, dynamic analysis can further analyze the behavior of the program by running it with different inputs and capturing the generated events at runtime. This technique is also more robust to code obfuscation, compared to static analysis. However, dynamic analysis is very resource-intensive to execute and may not be able to analyze the entire program, providing a less comprehensive view that could ignore malicious behaviors \cite{anderson2012improving}. In an attempt to benefit from both techniques, hybrid analysis is a solution that tries to combine the advantages of both static and dynamic analysis, while minimizing their weaknesses.

\subsubsection{Common Graph Structures for Malware Detection} \label{commonmalwaregraphs}

The various features extracted using the aforementioned analysis methods are frequently used to represent program semantics in the form of graphs. This practice has gained more and more interest due to the faculty of graphs to represent systems in a robust and intuitive way \cite{lee2010detecting,kruegel2006polymorphic,anderson2012improving}. The main graph structures  employed for malware detection are presented as follows:

\begin{controlflowgraph*}
Control flow graphs model all possible paths during the execution of a program, in an intra-procedural way. The nodes represent basic blocks, namely a sequence of instructions (e.g. assembly instructions) without any jumps. The jumps are characterized by the directed edges between the basic blocks, that represent the control flow of the program. When built from low-level assembly languages, CFGs have the faculty to be language-agnostic as they model the logic of a program without requiring language-specific instructions \cite{yan2019classifying}. Although other representations such as hexadecimal also possess this characteristic, CFGs provide a more intuitive way to model programs as graphs \cite{herath2022cfgexplainer,yan2019classifying}.
\end{controlflowgraph*}

\begin{fcg*} \label{cfg}
Function call graphs are a type of CFG that provide an inter-procedural view of the program, where nodes are functions and edges represent function calls from one function to another. Although FCGs offer a global view of function calls executed by the program, they generally lack the intra-procedural information that CFGs provide. To address this, some approaches can be employed by jointly using FCGs and CFGs, where embeddings from CFGs are integrated into the nodes of the FCGs, to capture both intra-procedural and inter-procedural semantic \cite{norouzian2021hybroid,xu2021hybrid,wang2021hierarchical}. In the case of Android malware analysis, a prevalent approach is to statically extract the API call sequences from the application and represent them using a FCG \cite{wei2018amandroid,li2022dmalnet,li2022intelligent,oliveira2019behavioral}.
\end{fcg*}


\begin{dependence*}
Program dependence graphs model both data and control dependencies in code, where nodes are instructions or statements and edges represent the data values and control conditions that must be fulfilled to execute the node’s operation \cite{ferrante1987program}. Scarcely used by current graph representation learning approaches, this graph structure may be promising to capture different conditional flows in the program \cite{xu2021android}.
\end{dependence*}

\begin{syscall*}
The system calls generated by the execution of a program can be captured via dynamic analysis and the communication with the system can be modeled with a graph, where nodes represent system calls and edges are the interactions \cite{john2020graph} between those calls. This representation offers a low-level view of system interactions that can also benefit the detection of malware.
\end{syscall*}

\begin{entities*}
During its execution, a program interacts with system entities such as processes, files, registry keys or network sockets. These interactions can be captured using sandbox tools like cuckoo \cite{cuckoo} for a deep analysis of the program's behaviors. Similarly as provenance graphs employed in host-based intrusion, these entities can be modeled as nodes in a graph, and the edges represent the operations between them \cite{hung2019malware,fan2020novel}. 
\end{entities*}

\begin{flows*}
The network activity generated by the program can also be monitored during its execution, and a network flow graph can be constructed with IP addresses and/or communication ports as nodes, and edges representing network flows. While some works solely rely on network traffic to detect malware activities \cite{busch2021nf, liu2023nt}, others enhance their detection capabilities by combining CFGs or FCGs with network data \cite{norouzian2021hybroid,xu2021hybrid}.
\end{flows*}

In this survey, we focus on the analysis of malware detection methods with graph representation learning for Android, Windows and Web platforms, since the vast majority of current state-of-the-art works are solely based on these platforms. For each of these platforms, we divided the state-of-the-art into sections based on the input graph structure. The literature review is summarized in Fig. \ref{fig:lit_surv}.

\definecolor{androidcolor}{HTML}{D6E8D5}
\definecolor{windowscolor}{HTML}{DAE8FC}
\definecolor{webcolor}{HTML}{E1D5E7}
\definecolor{myyellow}{HTML}{FFE6CC}

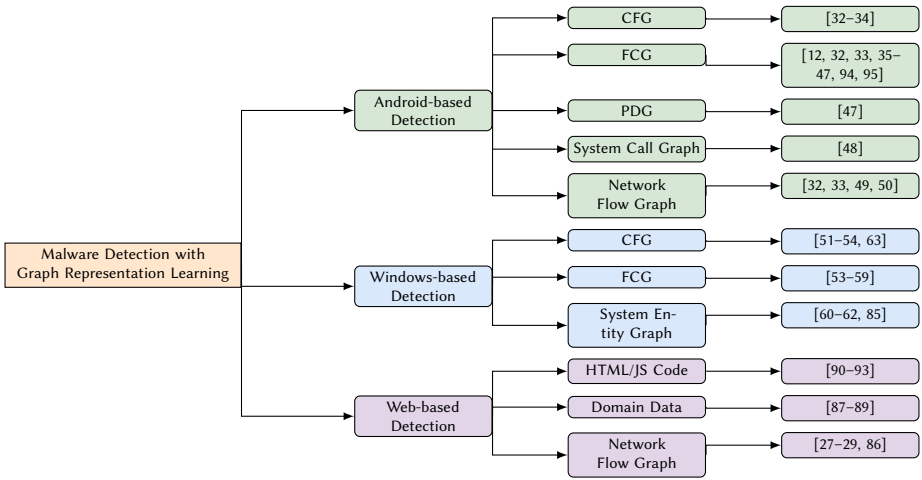
\begin{figure*}
    \centering

\smallskip


\tikzset{
    basic/.style  = {draw, text width=3cm, align=center, fill=myyellow, font=\sffamily, rectangle},
    root/.style   = {basic, rounded corners=2pt, thin, align=center, fill=green!30},
    androidnode/.style = {basic, thin, rounded corners=2pt, align=left, fill=androidcolor, text width=8em, align=center},
    windowsnode/.style = {basic, thin, rounded corners=2pt, align=left, fill=windowscolor, text width=8em, align=center},
    webnode/.style = {basic, thin, rounded corners=2pt, align=left, fill=webcolor, text width=8em, align=center},
    xnode/.style = {basic, thin, rounded corners=2pt, align=center, fill=blue!20,text width=2.5cm,},
    edge from parent/.style={draw=black, edge from parent fork right}

}
\tiny
\begin{forest} for tree={
    grow=east,
    growth parent anchor=west,
    parent anchor=east,
    child anchor=west,
    edge path={\noexpand\path[\forestoption{edge},->, >={latex}] 
         (!u.parent anchor) |- +(0pt,0pt) |-  (.child anchor) 
         \forestoption{edge label};}
}
[Malware Detection with \\Graph Representation Learning, basic,  l sep=15mm,
    [Web-based Detection, webnode,  l sep=10mm,
        [Network Flow Graph, webnode,  l sep=10mm,
            [\cite{zhou2020automating, carpenter2021detecting, zhao2020multi, li2022graphddos}, webnode]
         ],
        [Domain Data, webnode,  l sep=10mm,
            [\cite{he2019malicious, sun2020deepdom, zhang2021attributed}, webnode]
         ],
         [HTML/JS Code, webnode,  l sep=10mm,
            [\cite{ouyang2021phishing, secrypt22, liu2022graphxss, fang2022jstrong}, webnode]
         ]
        ]
    [Windows-based Detection , windowsnode,  l sep=10mm,
        [System Entity Graph, windowsnode,  l sep=10mm,
            [\cite{hung2019malware, wang2019heterogeneous, fan2020novel, liu2022fewm}, windowsnode]
         ],
        [FCG, windowsnode,  l sep=10mm,
            [\cite{jiang2018dlgraph, oliveira2019behavioral, zhang2020spectral, li2022intelligent, li2022dmalnet, wang2021hierarchical, ling2022malgraph}, windowsnode]
         ],
        [CFG, windowsnode,  l sep=10mm,
            [\cite{yan2019classifying, xu2021hawkeye, wang2021hierarchical, ling2022malgraph, herath2022cfgexplainer}, windowsnode]
         ],
         ]
    [Android-based Detection , androidnode,  l sep=10mm,
        [Network Flow Graph, androidnode,  l sep=10mm,
            [\cite{busch2021nf, liu2023nt, norouzian2021hybroid, xu2021hybrid}, androidnode]
         ],
         [System Call Graph, androidnode,  l sep=10mm,
            [\cite{john2020graph}, androidnode]
         ],
         [PDG, androidnode,  l sep=10mm,
            [\cite{xu2021android}, androidnode]
         ],
         [FCG, androidnode,  l sep=10mm,
            [\cite{zhu2018android, feng2020android, cai2021learning, xu2021detecting, vinayaka2021android, errica2021robust, catal2021malware, wu2022deepcatra, lo2022graph, gunduz2022malware, lu2022robust, yumlembam2022iot, hou2017hindroid, gao2021gdroid, hei2021hawk, xu2021android, xu2021hybrid, norouzian2021hybroid}, androidnode]
         ],
         [CFG, androidnode,  l sep=10mm,
            [\cite{norouzian2021hybroid,xu2021hybrid,fairbanks2021identifying}, androidnode]
         ]
         ]]
\end{forest}

    \caption{Categorization of current state-of-the-art papers in malware detection with graph representation learning. In this survey, we classify papers by platform and by input data structure. Android-based detection is presented in Section \ref{sec:android}, whereas Windows and Web detection are presented in Sections \ref{sec:windows} and \ref{sec:web}, respectively.}
    \label{fig:lit_surv}
\end{figure*}

\subsection{Methodology of Malware Detection with Graph Representation Learning}
\begin{figure*}[h]
\centerline{\includegraphics[width=\textwidth,keepaspectratio]{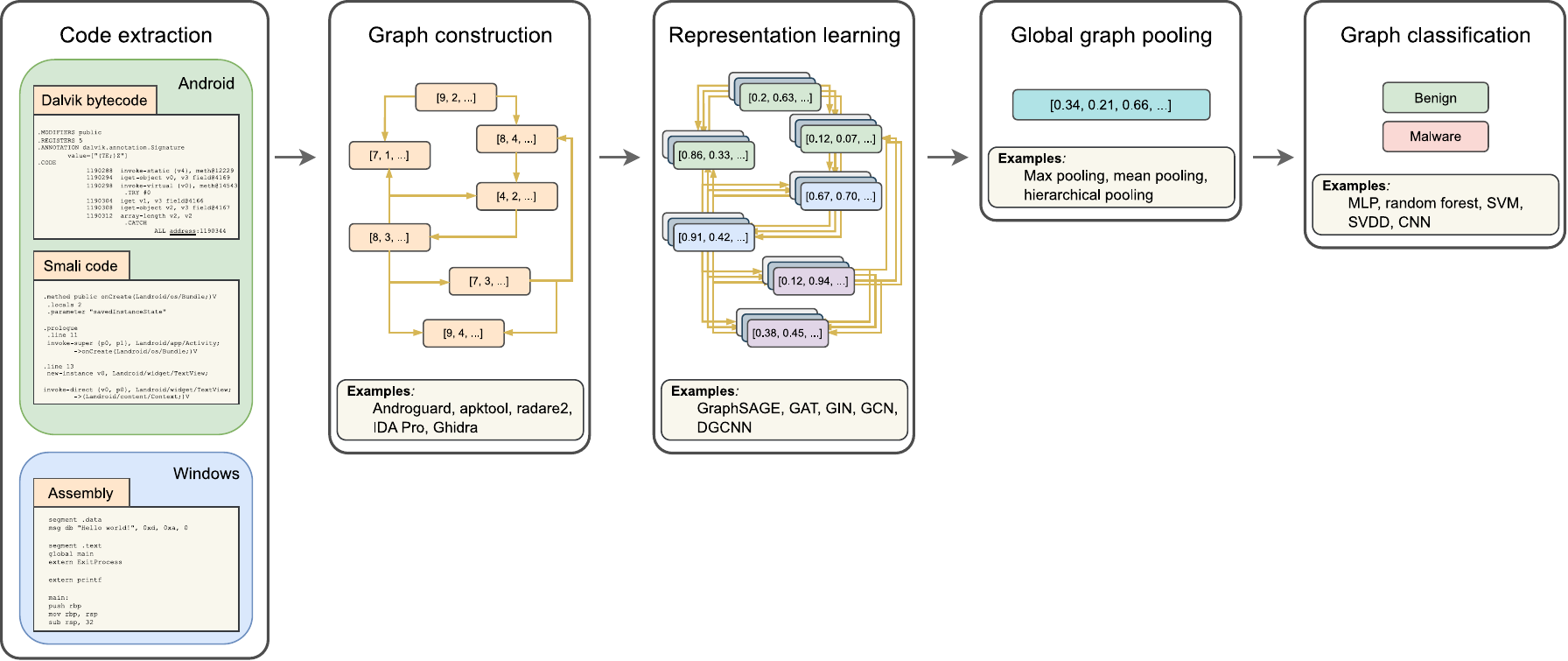}}
\caption{General architecture of malware detection from static code analysis based on graph representation learning.}
\label{fig:schema}
\end{figure*}

In literature, a majority of contributions rely on a similar sequence of operations to predict malware from source code with graph representation learning. In this section, we propose a general architecture, shown in Fig. \ref{fig:schema}, to summarize the process of malware detection from graph-represented source code on Android and Windows platform. 

The first step involves extracting code from the binary, which is usually disassembled to assembly language or decompiled to higher-level language. In the case of malware detection with dynamic analysis, this step assumes dynamic input features such as a stream of API calls or system entity interactions (see Section \ref{sec:windowsgraphs}). Subsequently, a graph builder is employed to transform the code into a graph-structured representation that preserves the program's semantics, as detailed in Section \ref{commonmalwaregraphs}. Typically, these first two steps are performed using reverse engineering tools listed in Table \ref{table:tools}. Optionally, the graph can be preprocessed and attributed with hand-crafted features, located on nodes or edges.
Then, graph representation learning techniques, such as GNNs, leverage the semantics of code to learn node embeddings, which capture the relationships and the role of internal instructions, functions, or API calls, depending on the input graph. These embeddings are commonly generated using well-known GNN variants, which have been discussed in Section \ref{section:gnn}. Other techniques employ word embedding techniques inspired from Natural Language Processing (NLP) to learn the meaning of opcode or API functions, enabling integration of the resulting embeddings into a global graph structure for GNNs to learn the structural properties. The majority of studies consider malware detection as a graph classification task, whereby node embeddings are transformed with a global pooling operation (or readout) to create a single fixed-size graph embedding vector that encapsulates all information of the graph. The final vector can then be classified using traditional ML or DL methods.

\definecolor{olivegreen}{HTML}{2E6930}

\begin{table}[]
\tiny
\caption{Common tools employed for data extraction and graph construction based on static or dynamic analysis.}
\centering
\begin{tabular}{@{}l|p{0.35\textwidth}@{}}\toprule
\textbf{Representation} & \textbf{Tools} \\
\midrule
CFG                     & \textcolor{olivegreen}{Androguard} \cite{androguard}, radare2 \cite{radare2}, IDA Pro \cite{ida}, Ghidra \cite{ghidra}  \\ 
FCG                     & \textcolor{olivegreen}{Androguard}, \textcolor{olivegreen}{Apktool} \cite{apktool}, \textcolor{olivegreen}{graph4apk} \cite{graph4apk}, \textcolor{olivegreen}{WALA} \cite{wala}, Angr \cite{angr}, radare2, IDA Pro, cuckoo \cite{cuckoo}, Ghidra  \\ 
PDG                     & \textcolor{olivegreen}{Androguard}, Ghidra \\
Syscalls                & strace \cite{strace}, SystemTap \cite{systemtap}, ltrace \cite{systemtap} \\ 
System entities         & cuckoo, Any.Run \cite{anyrun} \\ 
Network flows           & Argus \cite{argus}, Zeek \cite{zeek}, Splunk \cite{splunk}, Joy \cite{joy}  \\ 
\bottomrule
\end{tabular}
\begin{tablenotes}
  \setlength\labelsep{0pt}
    \footnotesize
    \parbox{\textwidth}{Green refers to tools specifically designed for Android APKs.}
\end{tablenotes}
\label{table:tools}
\end{table}

\section{Graph-based Android Malware Detection} \label{sec:android}
In this section, we present graph-based malware detection for Android platform, starting from the global methodology to build graphs from disassembled Android applications, to the review of existing works that leverage graph representation learning for the detection of malware.

\subsection{Android-based Graph Structures}
\begin{figure*}[h]
\centerline{\includegraphics[width=\textwidth,keepaspectratio]{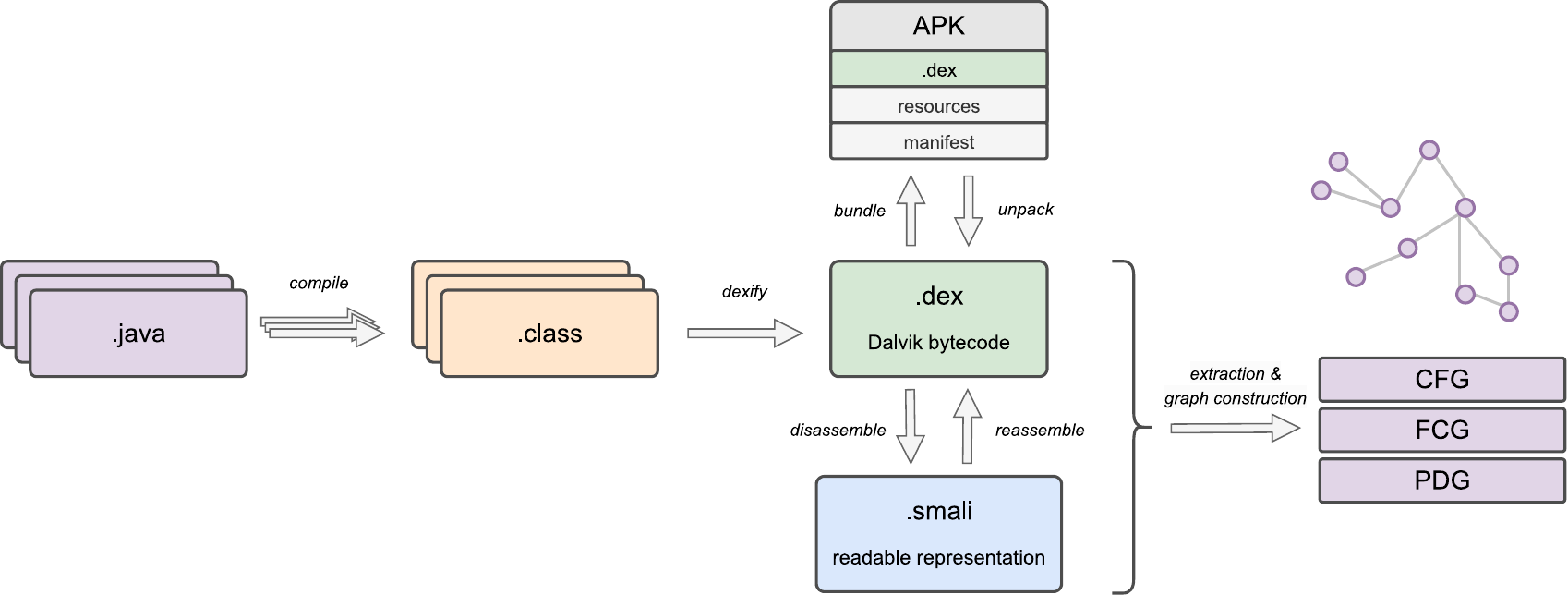}}
\caption{Android application compilation and disassembling process using static analysis. Java files are compiled into classes and are assembled into a single dex file. The APK is packed with the dex file, the manifest file and other resources. The dex bytecode can be disassembled into higher-level smali code and graph structures can be constructed from either dex or smali code depending on the use case requirements.}
\label{fig:android-disass}
\end{figure*}

Android applications are packed into APK files containing the source code, resources, manifest file and assets. After unzipping an APK, numerous features can be extracted to be used in downstream ML tasks. The manifest file provides the big picture of an app and contains its meta-information such as the required permissions to run it, hardware features and components. Resources such as images, videos or audio files are also available for further analysis. However, these data are inherently flat and do not provide enough structured information to build a graph. This is why most approaches leverage the actual source code to represent the logic of the app as a graph. As an APK is a production-ready app package, the code has already been compiled and assembled into a Dalvik bytecode format (.dex file). In practice, this bytecode can be disassembled into higher-level human-readable code (.smali files). Based on both code representations, control flow graphs (CFGs), function call graphs (FCGs), program dependence graphs (PDGs) and APIs can be extracted in a static way. This static extraction process is demonstrated in Fig. \ref{fig:android-disass}. Concerning dynamic analysis, API call and system call sequences can be recorded when running the app in a sandboxed environment. It is also possible to capture the network traffic by monitoring packets in a strategic area of a network.

\subsection{Android-based Approaches}
In this section, we review state-of-the-art Android-based papers, classified by types of graph, also summarized in Table \ref{table:androidsummary}.

\subsubsection{CFG Approaches for Android Malware Detection} \label{androidcfgbased}
CFGs offer a remarkable abstract representation of programs to detect malware. Hybroid \cite{norouzian2021hybroid} leverages this representation by extracting basic blocks from APKs. Three types of embeddings are then constructed from the code, to capture different semantics, namely opcode embedding, basic block embedding and CFG embedding, where each representation is associated to a level of abstraction. The semantic of opcodes (Dalvik instructions) and basic blocks (sequence of instructions) is computed using the NLP-based model word2vec with Skip-gram.
Precisely, Skip-gram learns embedding vectors for each basic block’s raw instructions, by using an opcode to predict its surrounding opcodes.
Indeed, the operands are not leveraged here as they are affected by the usage of Dalvik VM. For the basic block embeddings, they compute the weighted mean of the inner instructions’ opcode. These basic block embeddings then become the node embeddings from the point of view of a FCG and structure2vec \cite{dai2016discriminative} creates a final graph embedding vector for graph classification. In parallel, the network traffic generated by the app is also captured with dynamic analysis using Argus \cite{argus}. The packets are transformed into flows to summarize the communication between the Android device and the destination IP addresses, using various statistics. After a feature selection step, important features are combined with the FCG embeddings for downstream classification with gradient boosting on the CICAndMal2017 dataset, where the model demonstrates a F1-score of 97\% and beats other methods such as DREBIN \cite{arp2014drebin} and SVM \cite{guyon2002gene}.

On the other hand, hybrid-Falcon \cite{xu2021hybrid} transforms network flow data into 2D images on which a bi-directional LSTM captures flow representations from pixels. On the same dataset, the F1-score is further improved to 97.09\%.

The paper \cite{fairbanks2021identifying} provides a solution to locate MITRE ATT\&CK Tactics Techniques and Procedures (TTPs) detected from a subgraph of a CFG. Node representations are extracted with Inferential SIR-GN \cite{layne2021inferential} and the prediction is done using a random forest. To identify the subgraph responsible for the prediction, the authors rely on SHAP \cite{lundberg2017unified} to attribute for each input feature a value that indicates its relevance for the final output. TTPs are successfully detected with a F1-score of 92.7\%.

\begin{table*}[ht]
\tiny
\caption{Summary of Android-based malware detection approaches leveraging graph representation learning}
\centering
\begin{tabular}{@{}p{0.11\textwidth}p{0.05\textwidth}p{0.09\textwidth}p{0.12\textwidth}p{0.095\textwidth}p{0.19\textwidth}p{0.04\textwidth}p{0.15\textwidth}@{}}\toprule
\textbf{Data} & \textbf{Analysis} & \textbf{Graph type} & \textbf{Classification} & \textbf{Learning} & \textbf{Models} & \textbf{Year} & \textbf{Paper} \\

\midrule
\multirow{2}{*}{\textbf{CFG+FCG+Flows}}  & Hybrid & Attributed & Graph	& Supervised & Word2vec, structure2vec & 2021 & Hybroid \cite{norouzian2021hybroid} \\

  & Hybrid & Attributed & Graph & Supervised & Bi-LSTM, word2vec, structure2vec & 2021 & hybrid-Falcon \cite{xu2021hybrid} \\
 
 \cmidrule{1-8}
 \multirow{1}{*}{\textbf{CFG}} & Hybrid & Attributed & Graph, Subgraph & Supervised & Inferential SIR\_GN, RF & 2021 & Fairbanks et al. \cite{fairbanks2021identifying} \\
 
\cmidrule{1-8}
 \multirow{12}{*}{\textbf{FCG}}  & \multirow{12}{*}{Static} & \multirow{12}{*}{Attributed} & \multirow{12}{*}{Graph} & \multirow{12}{*}{Supervised} & GCN & 2018 & CG-GCN \cite{zhu2018android} \\

 &  &  &  &	 &	GNN & 2021 &	CGDroid \cite{feng2020android} \\

 &  &  &  &  & GCN, CBOW & 2021 & Cai et al. \cite{cai2021learning} \\

 &	 &	 &	 &	 & GNN, Skip-gram & 2021 & Xu et al. \cite{xu2021detecting} \\

 &	 &	 &	 &	 &	GraphSAGE & 2021 &	Vinayaka et al. \cite{vinayaka2021android} \\

 &	 &	 &	 &	 &	CGMM & 2021 &	Errica et al. \cite{errica2021robust} \\

 &	 &	 &	 &	 &	GAT, node2vec & 2021 &	Catal et al. \cite{catal2021malware} \\

 &	 &	 &	 &	 &	GNN, Bi-LSTM, TF-IDF & 2022 &	DeepCatra \cite{wu2022deepcatra} \\

 &	 &	 &	 &	 &	GCN, GraphSAGE, GIN & 2022 &	Lo et al. \cite{lo2022graph} \\
 
 &	 &	 &	 &	 &	VGAE, word2vec & 2022 & Gunduz et al. \cite{gunduz2022malware} \\

 &	 &	 &	 &	 &	GCN & 2022 &	Lu et al. \cite{lu2022robust} \\

 &	 &	 &	 &	 &	GraphSAGE, VGAE & 2022 &	Yumlembam et al. \cite{yumlembam2022iot} \\

\cmidrule{1-8}
\multirow{3}{*}{\textbf{App-API FCG}} & \multirow{3}{*}{Static} & \multirow{3}{*}{Heterogeneous} & \multirow{3}{*}{Node} & \multirow{3}{*}{Supervised} & Multi-kernel model, Meta-path & 2017 & HinDroid \cite{hou2017hindroid} \\

&  &  &  &  & GCN, Skip-gram & 2021 & GDroid \cite{gao2021gdroid} \\

&  &  &  &  & Custom HAN, Meta-path & 2021 & Hawk \cite{hei2021hawk} \\

\cmidrule{1-8}
\multirow{1}{*}{\textbf{PDG+FCG}}  & Static & Attributed & Graph & Supervised & structure2vec, word2vec, SIF & 2021 & Android-COCO \cite{xu2021android} \\

\cmidrule{1-8}
\multirow{1}{*}{\textbf{Syscall Graph}}  & Dynamic & Attributed & Graph & Supervised & GCN & 2020 & John et al. \cite{john2020graph} \\

\cmidrule{1-8}
\multirow{2}{*}{\textbf{Flow Graph}}  & Hybrid & Attributed & Graph & Supervised, Unsupervised & GNN, GAE, Residual connections, Deep SVDD	& 2021 & NF-GNN \cite{busch2021nf} \\

& Hybrid & Attributed & Graph & Supervised & MPNN, GRU & 2023 & NT-GNN \cite{liu2023nt} \\

\bottomrule
\end{tabular}
\begin{tablenotes}
  \setlength\labelsep{0pt}
    \footnotesize
    \textbf{Data} represents the data type taken as input by the models; \textbf{Analysis} refers to the analysis method that is leveraged to extract features (e.g. extracting a CFG from smali code is static, capturing network traffic from a running app is dynamic, whereas leveraging both results in a hybrid analysis); \textbf{Graph type} designates one of the graphs introduced in Section \ref{sec:graphs}, here we characterize a graph as attributed if a node or an edge is attributed either with hand-crafted features, raw features (e.g. raw instructions, function names) or embeddings (e.g. word embedding of a function), whereas a heterogeneous graph deals with multiple types and possibly different attributes; \textbf{Classification} designates the final object to classify (i.e. the classification task); \textbf{Learning} is the learning method used to train the models, whereas \textbf{Models} refer to the models on which the paper is inspired; \textbf{Paper} and \textbf{Year} identify the work and its publication year.
\end{tablenotes}
\label{table:androidsummary}
\end{table*}

\subsubsection{FCG Approaches for Android Malware Detection}
The semantic information captured by FCGs in programs makes this data structure a predominant choice in graph-based malware detection. For instance, a FCG is constructed from Smali code in work \cite{zhu2018android}, where each node is attributed with function attributes such as the method type (system API, third-party API, etc.) and the requested permissions (required permissions for the execution of the function). The graph embeddings are then trained in a supervised way using the GCN propagation rule, presented in Eq. \ref{eq:gcn1} and \ref{eq:gcn2}. 

\begin{align*}
H^{(k)} &= \sigma\left(\mathcal{A} H^{(k-1)} \textbf{W}^{(k)}\right) \addtag \label{eq:gcn1}
\end{align*}
where $H^{(k)}$ represents the node embedding matrix at layer $k$ with $H^0=X$, the feature matrix. $\sigma$ represents an activation function, $\textbf{W}^{(k)}$ is a trainable weight matrix and $\mathcal{A}$ designates the normalized adjacency matrix with self-loops described below.

\begin{align*}
\mathcal{A} &= \tilde{D}^{-\frac{1}{2}}(A+I)\tilde{D}^{-\frac{1}{2}} \addtag \label{eq:gcn2}
\end{align*}
where $I$ the identity matrix and $\tilde{D}$ is the degree matrix of adjacency matrix with self-loops $A+I$. The Drebin dataset is used for final evaluation with a F1-score of 99.68\%.

In CGdroid \cite{feng2020android}, multiple node features are extracted from the disassembled methods in order to build a FCG that captures the semantic of functions. Indeed, each node is mapped to a vector of hand-designed features such as the number of string constants, the number of call and jump instructions, the associated permissions, etc. A GNN computes graph embeddings and a MLP is used for downstream graph classification on Drebin and Androzoo \cite{allix2016androzoo} datasets, where a baseline is outperformed by 8\% in F1-score.

Word embedding techniques are employed in \cite{cai2021learning} to consider functions similarly as words and learn the meaning of functions. The embeddings are then assigned as attributes to each corresponding function node in a FCG, and a GCN is used as graph learning method. The proposed method achieved 99.65\% F1-score with random forest classifier on a private dataset.

Similarly, authors in \cite{xu2021detecting} leverage word embedding to transform Android opcodes from text to vectors using Skip-gram. In the same way as \cite{cai2021learning}, the embeddings are used as nodes in a FCG and this graph is fed into a GNN to compute a fixed-size graph embeddings vector. A 2-layer MLP and softmax are used as last layers of the architecture for graph classification, with an average accuracy of 99.6\%. 

In the reference \cite{vinayaka2021android}, FCGs are extracted from APKs using Androguard and each node stores attributes related to the structural meaning of the node in the graph (e.g. node degree) or features extracted from the actual disassembled functions (e.g. method attributes, method opcodes’ summary). Using these previous features, GCN, GraphSAGE, GAT and TAGCN \cite{du2017topology} are benchmarked together, with a better performance achieved with GraphSAGE. First, each node $i$ uniformly selects  a fixed-size set of neighbors, denoted $\mathcal{N}(i)$. Neighbors are then aggregated using a mean aggregation function such as:

\begin{align*}
     & \mathbf{h}_{\mathcal{N}(i)}^{(l+1)}=\operatorname{aggregate}\left(\left\{\mathbf{h}_j^{(l)}, \forall j \in \mathcal{N}(i)\right\}\right) \addtag
\end{align*}
where $\mathbf{h}_{\mathcal{N}(i)}^{(l+1)}$ is the embedding of node $i$ at layer $l+1$ and $h_j^{(l)}$ denotes the embedding of a neighbor node $j$ at layer $l$. The embedding of $i$ at previous layer is then concatenated with the aggregated representation and then learned by a neural network.
\begin{align*}
\mathbf{h}_i^{(l+1)}=\sigma\left(\mathbf{W}^{(l)} \cdot \left[\mathbf{h}_i^{(l)}, \mathbf{h}_{\mathcal{N}(i)}^{(l+1)}\right]\right) \addtag
\end{align*}
where $[,]$ represents the concatenation operation. The embedding is finally normalized:
\begin{align*}
\mathbf{h}_i^{(l+1)}=\frac{\mathbf{h}_i^{(l+1)}}{||\mathbf{h}_i^{(l+1)}||_2} \addtag 
\end{align*}
Malware Android apps from CICMalDroid2020 dataset are used for evaluation, with a best F1-score of 92.23\%.

For the detection of obfuscated malware, the paper \cite{errica2021robust} leverages a call graph where nodes are attributed with nodes' out-degree only. The Contextual Graph Markov Model (CGMM) \cite{bacciu2018contextual} is used to learn the embeddings that are then classified using a standard feed-forward network, achieving a macro F1-score of 97.2\%.

In the study \cite{catal2021malware}, the node embeddings of an API call graph are computed using node2vec and aggregated with a GAT model. 
The authors explain that the use of node2vec as feature extraction method is justified by a 3\% increase in F-score compared to traditional graph centrality features. Using the attention aggregation from GAT as a final step, the proposed solution reaches 94.1\% in F-score.

API call traces and opcode features are further exploited in DeepCatra \cite{wu2022deepcatra}, where Term Frequency-Inverse Document Frequency (TF-IDF) is used to identify critical Android APIs from call traces. The prior knowledge required to detect critical APIs is extracted from popular codebases available in online repositories such as CVE \cite{cve} and Exploit-DB \cite{exploitdb}. Then, call graphs are generated from apps with Wala \cite{wala} by also considering the knowledge of previously identified  critical APIs. Finally, a custom GNN and a bi-directional LSTM are trained in an end-to-end manner, to respectively capture the graph topology and temporal features from call traces. The output vectors produced by both models are then merged with a fully connected layer and a softmax layer is used for binary classification. The proposed model is evaluated against \cite{john2020graph} and other CNN-, GCN- and LSTM-based baselines. Compared to the baselines, DeepCatra achieves best results with 95.83\% F1-score and further improves false positive and false negative rates.

In the paper \cite{lo2022graph}, the authors build an enhanced FCG, where node attributes are graph centrality indicators based on graph nodes' importance. Precisely, PageRank \cite{page1999pagerank}, in/out degree and node betweenness centralities are used as node features. The FCG is fed into a GCN, a GraphSAGE and a GIN model for comparison, and all leverage jumping knowledge \cite{xu2018representation}, a technique to overcome over-smoothing in GNN architectures by introducing jump connections in the neural network. GraphSAGE outperforms the compared baselines by an important gap on the multi-class classification task, with a F1-score respectively of 94\% and 97\% on Malnet-Tiny \cite{freitas2020large} and Drebin \cite{arp2014drebin} datasets.

Gunduz et al. \cite{gunduz2022malware} compute node-level embeddings from sensitive API function calls using word2vec and leverage a Variational Graph Auto-Encoder (VGAE) \cite{kipf2016variational} to learn a reduced representation of embeddings that is used for downstream classification, with a F-measure of 93.4\%.

For the detection of obfuscated malware from call graphs, the paper \cite{lu2022robust} leverages a GCN with subgraphs along with a denoising method. The method is evaluated on a private dataset made of samples from VirusShare and AndroZoo, whereas Proguard \cite{proguard} is used for code obfuscation.

A more general approach is proposed by Yumlembam et al. \cite{yumlembam2022iot}, who consider each Android app as a local graph, where nodes are APIs and an edge exists between two APIs if they co-exist in a same code block (i.e. a code segment from a smali file, located between \textit{.method} and \textit{.endmethod}). A global graph then represents the connections among applications with co-occuring APIs. Multiple variants of the model are compared, using different features. One variant considers attributed nodes with  5 centrality indicators: degree, betweenness, closeness, eigenvector and PageRank. Another variant considers the permissions and intents from the manifest file. After benchmarking multiple models, the best combination is to calculate graph embeddings with GraphSAGE and to concatenate the vector with the permissions and intents features. The resulting vector is then passed into a classifier trained in supervised fashion, where a traditional CNN achieves best performance. In order to test the robustness of GNN-based malware detection models, the authors also provide a generative model inspired from VGAE, which can generate adversarial API graphs to fool the predictive model. The proposed methods are finally compared to many state-of-the-art techniques and  respectively achieve 98.33\% and 98.68\% accuracy in Drebin and CICMaldroid datasets.

Other works model the interactions between APKs and API calls as a global view. In this paper, we describe this representation as an App-API graph, where a node is an Android app or an API call and an edge is a relation between two endpoint nodes such as two apps containing a same API, or two APIs coming from a same package. In literature, this type of structure is mostly represented as a large Heterogeneous Information Network (HIN) \cite{shi2016survey} (i.e. an heterogeneous graph), where the goal is to classify malware app nodes. In HinDroid \cite{hou2017hindroid,hou2018make}, an App-API graph models the interactions between Android apps and APIs, as a HIN. A node is either an app or an API call, whereas an edge is one among multiple relations representing whether extracted API calls belong to the same code block, or if they are with the same package name, or use the same invoke method.
Semantic is extracted from the heterogeneous graph using meta-paths. A meta-path is a path composed of a series of different node and edge types that captures a particular semantic in the graph. Because of their heterogeneous composition and significant semantic extraction abilities, meta-paths are frequently used in heterogeneous graphs.
These meta-paths constructed by the traversal of graph are leveraged by a multi-kernel SVM to determine a weight for each meta-path. These weights are then considered for downstream app’s node classification with a final F1-score of 98.84\%. 

 Gdroid \cite{gao2021gdroid} is another technique that extracts a graph from API co-occurrence in APKs to build an App-API graph. The Skip-gram model first encodes APIs while preserving context information, with the objective to obtain similar embeddings for APIs with similar usages. On top of the graph, a GCN propagates the information and learns node embeddings that are leveraged for downstream node classification with 98.99\% accuracy. 

 In Hawk \cite{hei2021hawk}, more than 180k APKs are extracted to build a large heterogeneous App-API graph that also models relations such as App-Permission, App-Class or App-Interface. Meta-paths along with meta-graphs are extracted from the graph to capture semantics. More precisely, two models are built for in-sample and out-sample nodes. The former is based on a custom heterogeneous GAT that fully leverages meta-structures to capture embeddings whereas the latter utilizes these embeddings in an incremental setting to quickly learn new embeddings without requiring re-learning. Hawk was evaluated against many baselines and outperforms them by a large gap on both in-sample and out-sample malware detection.

\subsubsection{PDG Approaches for Android Malware Detection}
Program Dependence Graphs have been widely used in optimization tasks due to their faculty to model data and control flow from programs \cite{ferrante1987program}. For similar reasons, this structure is also employed in malware detection but remains little used with graph representation learning.

Android-COCO \cite{xu2021android} leverages the native code of dynamic libraries (.so files) along with the Android bytecode (.dex files) to construct a PDG for each app. Structure2vec computes the graph embeddings, that are then passed into a MLP for graph classification. For a more accurate prediction, a FCG is created, on which graph embeddings are also computed (similarly than Hybroid \cite{norouzian2021hybroid}). The predictions of both graphs are finally combined using an ensemble algorithm and a 99.88\% F1-score is reached on samples from Drebin, AMD \cite{wei2017deep} and Androzoo datasets.

\subsubsection{System Call Approaches for Android Malware Detection}
System calls provide a low-level view of system interactions, able to model attacks patterns.
The authors in \cite{john2020graph} rely on dynamic analysis to record the system calls generated by the activity of a running APK to detect malware behaviors. Each node in the graph is one among 26 selected system calls and is summarized by 4 centrality indicators as node features: Katz, Betweennes, Closeness and PageRank centralities. Edges represent interactions between those system calls while the app is running. A GCN and a pooling layer compute graph embeddings and a fully-connected layer along with a softmax activation are used for graph classification. Their implementation achieves 92.3\% accuracy and similar true positive rate as SVM but significantly outperforms  all other methods regarding the false positive rate.
As for the PDG, system call graphs are still scarcely used in current graph representation learning literature.

\subsubsection{Network Flow Approaches for Android Malware Detection}
NF-GNN \cite{busch2021nf} proposes to detect Android malware using network flows constructed from pcap network captures. Only IP addresses are used to build the graph structure, the source and destination ports are not leveraged here. They propose a custom GNN model with a propagation function that considers both edge and node features. A MLP is first used on edge features and endpoint nodes aggregate the edges using the concatenation operation along with a residual connection \cite{he2016deep}. Then, three downstream methods are proposed for classification based on graph embeddings: a graph classifier (supervised), a custom graph autoencoder (unsupervised), and a one-class neural network (unsupervised). The graph classifier consists in adding a pooling layer and a dense layer with softmax activation. The GAE method uses as encoder the GNN described previously and as decoder a custom model that tries to reconstruct the original edge features from the compact node representations produced by the GNN encoder. Finally, the one-class network alternative is described as a pooling layer followed by a Deep Support Vector Data Description (SVDD) \cite{ruff2018deep} model. The three variants have been compared to 7 supervised and unsupervised baseline methods on the CICAndMal2017 dataset and all methods outperform compared baselines by an important gap.

Similarly, NT-GNN \cite{liu2023nt} monitors the traffic produced by running APKs and converts the packets into flows using CICFlowMeter-V3. Communication ports are don’t considered and only IP addresses from flows are leveraged to build the graph. A model inspired from the MPNN is used for message-passing between nodes in the network graph and the classification is performed after applying a readout of the computed node embeddings. Node representations are updated by passing the previous representation with the new aggregated representation from neighbors into a GRU and the model is trained using cross-entropy loss. A 97\% F1-score is reached on both CICAndMal2017 and AAGM datasets.

Other flow-based works are presented in \cite{norouzian2021hybroid,xu2021hybrid}, where the authors leverage network flows in combination with CFGs and FCGs to further improve the prediction capacities of the model (see Section \ref{androidcfgbased}).

\subsection{Android Malware Datasets}
In this section, we present Android datasets employed for graph-based malware detection tasks. A summary of the datasets used in previous studies is available in Table \ref{table:androiddatasets}. Based on the current information provided from the respective websites of AMD, Malgenome and PRAGuard datasets, the release of these datasets has stopped for maintenance reasons and are not further described in this paper.

\begin{table*}[htb]
\tiny
\caption{Datasets employed in Android malware detection studies.}
\centering
\begin{tabular}{@{}p{0.15\textwidth}p{0.3\textwidth}@{}p{0.15\textwidth}}\toprule
\textbf{Paper} & \textbf{Datasets} & \textbf{Performance} \\

\midrule

Hybroid \cite{norouzian2021hybroid} & CICAndMal2017 & 97\% F1 \\
hybrid-Falcon \cite{xu2021hybrid} & CICAndMal2017,AndroZoo & 97.09\% F1 \\
Fairbanks et al. \cite{fairbanks2021identifying} & VirusTotal & 92.7\% F1 \\
CG-GCN \cite{zhu2018android} & Drebin+Apkpure+HKUST & 99.68\% F1 \\
CGDroid \cite{feng2020android} & Drebin+AndroZoo & \textasciitilde 99\% F1 \\
Cai et al. \cite{cai2021learning} & AndroZoo+VirusShare & 99.65\% F1 \\
Xu et al. \cite{xu2021detecting} & Drebin+AMD+PRAGuard+AndroZoo & 99.6\% acc \\
Vinayaka et al. \cite{vinayaka2021android} & CICMalDroid2020,AndroZoo & 92.23\% F1 \\
Errica et al. \cite{errica2021robust} & AMD,Google Play Store & 97.2\% macro F1 \\
Catal et al. \cite{catal2021malware} & CICMalDroid2020+ISCX-AndroidBot-2015 & 94.1\% F1 \\
DeepCatra \cite{wu2022deepcatra} & \parbox{0.25\textwidth}{Drebin+DroidAnalytics+VirusShare +CICInvesAndMal2019+AndroZoo} & 95.83\% F1 \\
Lo et al. \cite{lo2022graph} & MalNet-Tiny,Drebin & 94\%, 97\% F1 \\
Gunduz et al. \cite{gunduz2022malware} & ISCX-AndroidBot-2015+CICMalDroid2020 & 93.4\% F1 \\
Lu et al. \cite{lu2022robust} & VirusShare+AndroZoo+Google Play Store & \textasciitilde 63\%-95\% F1 \\
Yumlembam et al. \cite{yumlembam2022iot} & Drebin,CICMalDroid2020 & 98.33\%, 98.68\% acc \\
HinDroid \cite{hou2017hindroid} & Private & 98.84\% F1 \\
GDroid \cite{gao2021gdroid} & AMD,Google Play Store & 98.99\% acc \\
Hawk \cite{hei2021hawk} & \parbox{0.25\textwidth}{CICAndMal2017+VirusShare+AndroZoo +Google Play Store} & >96\% F1 \\
Android-COCO \cite{xu2021android} & Drebin+AMD+AndroZoo & 99.88\% F1 \\
John et al. \cite{john2020graph} & Drebin+AMD+Malgenome & 92.3\% acc \\
NF-GNN \cite{busch2021nf} & CICAndMal2017 & 96.75\% AUC, \textasciitilde 99\% F1 \\
NT-GNN \cite{liu2023nt}	& CICAndMal2017,AAGM & 97\%, 97\% F1 \\

\bottomrule
\end{tabular}
\begin{tablenotes}
  \setlength\labelsep{0pt}
    \footnotesize
  For each \textbf{Paper}, we provide the list of datasets along with the performance of the model. \textbf{Datasets} separated by a "+" refer to a global dataset built from the assembling of each mentionned dataset, whereas the use of a "," means that the authors have conducted experiments on separated datasets. If multiple comma-separated datasets are present in Datasets, and only one metric is assigned in Performance, then this metric refers to the performance of the first dataset. Otherwise, each dataset is assigned to a performance metric. If the number of metrics in Performance is greater than the number of datasets, then multiple variants of the models are proposed and we suggest to refer to the original paper for further information. In this paper, \textbf{Performance} metrics are defined such that the accuracy denoted as "acc"=$(TP+TN)/(TP+TN+FP+FN)$, where $TP$, $TN$, $FP$, $FN$ refer to true positive, true negative, false positive and false negative, respectively. "F1"=$2\times$Precision$\times$Recall$/($Precision $+$ Recall$)$ where Precision$=T P /(T P+F P)$ and Recall$=T P /(T P+$ $F N)$. "AUC" refers to the  Area Under the Receiver Operating Characteristic Curve.
\end{tablenotes}
\label{table:androiddatasets}
\end{table*}

\begin{cicandmal*}
An Android malware dataset developed by the Canadian Institute of Cybersecurity (CIC). It comprises 10,854 APK files published between 2015 and 2017 on Google Play Store. The dataset consists of 6,500 benign apps and 4,354 malware divided into Benign, Adware, Ransomware, SMS and Riskware classes. For each scenario, network packets are also collected and transformed into flows using CICFlowMeter \cite{cicflowmeter}. This tool generates, for each flow, 80 features based on statistics from the packets contained within the flow. 
\end{cicandmal*}

\begin{cicmaldroid*}
This dataset was also made public by the  Canadian Institute for Cybersecurity. It is composed of 17,341 APK samples collected during one year in 2018. Malware examples are divided into 5 classes: Benign, Adware, Banking, SMS and Riskware. Along with the APK files that can be used for classification tasks, three kinds of features are also provided for each sample: statically extracted features (e.g. intents, permissions and services), dynamically observed behaviors (e.g. system calls, binder calls, composite behaviors) and network traffic in pcap format. Features are available from CSV files, ranging from 139 to 50,621 files depending on the APK.
\end{cicmaldroid*}

\begin{androzoo*}
A collection of Android apps provided by the University of Luxembourg. In 2022, the dataset contains more than 21M APKs, mostly including benign apps from Google Play Store but also malware from VirusShare. Many other APK stores are fetched to continually update the collection. For each APK file, 9 features are collected such as the sha256 hash, the app compilation date or the size of the .dex file. AndroZoo is often used in combination with other APK malware datasets to obtain a balanced number of benign samples.
\end{androzoo*}

\begin{drebin*}
Made available by the MobileSandbox project, this dataset also provides malware Android apps. A total of 5,560 APKs divided into 179 malware families were collected between August 2010 and October 2012. Considering the important variety of classes, most multi-class classification papers use the top-k classes from the dataset by sorting malware based on the number of samples per class. Otherwise, all examples can be used for binary classification. Each APK is summarized by 10 features such as permissions, intents and providers. This dataset does not contain any benign example so an additional dataset such as AndroZoo should be used to complete the dataset with benignware examples. 
\end{drebin*}

\begin{malnet*}
A large dataset containing FCGs extracted from AndroZoo APK files. According to the original paper from 2021, it was at this time the largest database for graph representation learning with 1,262,024 graphs, averaging over 15k nodes and 35k edges per graph, divided among 47 types. GNNs have been applied to this dataset in the original paper \cite{freitas2020large}, where baselines such as GCN, GIN or Feather are benchmarked together. Moreover, FCGs have demonstrated promising results when combined with representation learning techniques, in trying to overcome the polymorphic nature of malware \cite{gascon2013structural,jiang2018dlgraph}. For smaller experiments, MalNet-Tiny is a subset of MalNet composed of 5,000 graphs of at most 5k nodes and balanced in 5 types. The authors also released MalNet Explorer \cite{malnetexplorer}, a useful web interface to explore the graph structures of malware from the dataset.
\end{malnet*}

\section{Graph-based Windows Malware Detection} \label{sec:windows}
The increasing level of sophistication of malware on Windows platform along with the widespread usage of this operating system worldwide has become a major concern to preserve the safety of many users. After the success of graph representation learning in many classification tasks, its application to Windows-based malware detection has become obvious. As for Android malware detection, in this section, we present a global methodology to model Windows binaries as graphs and we perform a literature review of current approaches involving graph learning algorithms.

\subsection{Windows-based Program Graph Structures} \label{sec:windowsgraphs}
\begin{figure*}[h]
\centerline{\includegraphics[width=\textwidth,keepaspectratio]{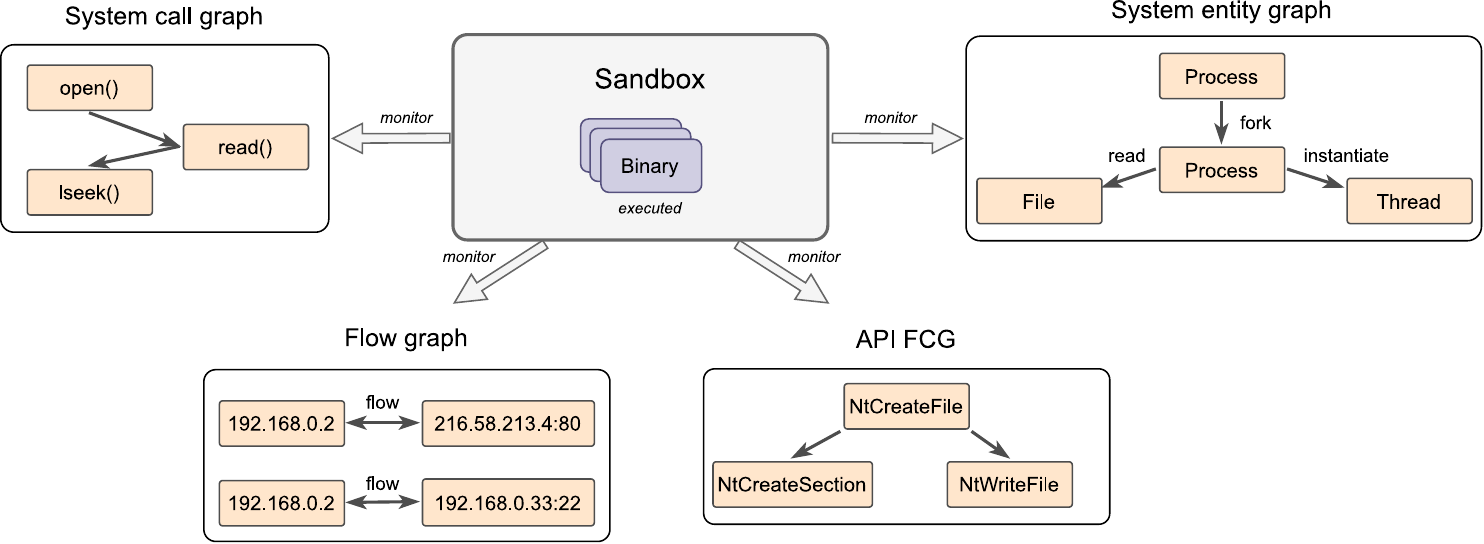}}
\caption{Extraction of graph structures from a running binary using dynamic analysis. The file is safely executed in a sandbox environment and all system- and network-level events are monitored for downstream graph construction.}
\label{fig:windowssandbox}
\end{figure*}

In Windows, executable files are encapsulated following the portable executable (PE) file format. File extensions such as .exe, .dll and .sys are all PEs with different roles. Dynamic-link libraries (DLL) and system (SYS) files are both libraries of functions that are loaded into memory and used by other programs. The former is intended for a general function sharing purpose, whereas the latter is intended for a more specific use related to device drivers and hardware configurations by the system \cite{hahn2014robust}. The EXE file is the one that is actually executed and that communicates with function libraries. Similarly as for Android APKs, PE files can be analyzed statically to extract source code employed in downstream graph structures such as CFGs, FCGs and PDGs. PEs are compiled code, meaning that the binary has to be first disassembled or decompiled to obtain an appropriate human-readable representation like assembly. A number of works also leverage dynamic analysis to detect Windows malware, where the executable binaries are run into a sandbox to monitor system entity interactions, system calls, network flows and live API calls. This process is described in Fig. \ref{fig:windowssandbox}.

\subsection{Windows-based Approaches}
This section reviews the approaches employed in the Windows-based papers presented in Table \ref{table:windows}.

\begin{table*}[ht]
\tiny
\caption{Summary of Windows-based malware detection approaches leveraging graph representation learning.}
\centering
\begin{tabular}{@{}p{0.06\textwidth}p{0.05\textwidth}p{0.11\textwidth}p{0.1\textwidth}p{0.08\textwidth}p{0.21\textwidth}p{0.04\textwidth}p{0.15\textwidth}@{}}\toprule
\textbf{Data} &  \textbf{Analysis} & \textbf{Graph type} & \textbf{Classification} & \textbf{Learning} & \textbf{Models} & \textbf{Year} & \textbf{Paper} \\
\midrule
\multirow{2}{*}{\textbf{CFG}} & Static & Attributed & Graph & Supervised & DGCNN & 2019 & MAGIC \cite{yan2019classifying} \\

& Hybrid & Attributed & Graph & Supervised & GNN, word2vec & 2021 & HawkEye \cite{xu2021hawkeye} \\

 \cmidrule{1-8}
 \multirow{2}{*}{\textbf{CFG+FCG}}  & Static & Attributed & Graph & Supervised & GAT, Random Walk, BERT & 2021 & Wang et al. \cite{wang2021hierarchical} \\

 & Static & Attributed & Graph & Supervised & GraphSAGE & 2022 & MalGraph \cite{ling2022malgraph} \\
 
 \cmidrule{1-8}
 \multirow{5}{*}{\textbf{FCG}} & Static & Attributed & Graph & Supervised & node2vec, SDA & 2019 & DLGraph \cite{jiang2018dlgraph} \\
 
  & \multirow{4}{*}{Dynamic} & \multirow{4}{*}{Attributed} & \multirow{4}{*}{Graph} & \multirow{4}{*}{Supervised}  & DGCNN & 2019 & Oliveira et al. \cite{oliveira2019behavioral} \\

  &  &  &  &  & GCN & 2021 & SDGNet \cite{zhang2020spectral} \\
 
 &  &  &  &  & GCN, Markov chain & 2021 & Li et al. \cite{li2022intelligent} \\
 
 &  &  &  &  & GAT, GIN, Word2vec & 2022 & DMalNet \cite{li2022dmalnet} \\
 
 \cmidrule{1-8}
 \multirow{4}{*}{\textbf{Entities}} & Dynamic & Attributed & Graph & Supervised & GCN & 2019 & MeQDFG \cite{hung2019malware} \\

 & Dynamic & Heterogeneous & Graph & Semi-supervised & Heterogeneous GNN, Meta-path, Attention, Siamese Network	& 2019 & MatchGNet \cite{wang2019heterogeneous} \\
 
 & Dynamic & Heterogeneous, Attributed & Node & Supervised & GraphSAGE, Meta-path & 2021 & MalSage \cite{fan2020novel} \\

 & Dynamic & Heterogeneous & Graph & Self-supervised & GAT, Meta-path, Contrastive learning &  2022	& FewM-HGCL \cite{liu2022fewm} \\
 
\bottomrule
\end{tabular}
\begin{tablenotes}
  \setlength\labelsep{0pt}
    \footnotesize
    
\end{tablenotes}
\label{table:windows}
\end{table*}

\subsubsection{CFG Approaches for Windows Malware Detection}
In MAGIC \cite{yan2019classifying}, assembly code is extracted from PE files and converted into CFG, where nodes are basic blocks composed of multiple assembly instructions, and edges represent the program flow along these basic blocks, as explained in Section \ref{commonmalwaregraphs}. Instead of using a standard GCN that was initially made for node classification, the authors prefer to leverage a Deep Graph Convolutional Neural Network (DGCNN) \cite{zhang2018end}, which is especially designed for graph classification. The proposed DGCNN leverages adaptive max pooling and replaces the original Conv1D layer with a custom layer that considers graph embedding idea. The training procedure of this model minimizes the mean negative logarithmic loss in an end-to-end manner. The authors trained the model for malware classification on two private CFG-based datasets: MSKCFG (inspired by \cite{ronen2018microsoft}) and YANCFG (inspired by \cite{yan2015sensitive}). These datasets were not made publicly available, but the procedure to generate the CFGs is provided in the paper. The model was also evaluated on the Microsoft Malware Classification Challenge dataset \cite{ronen2018microsoft} and reaches 99.25\% accuracy.

A cross-platform approach is proposed in HawkEye \cite{xu2021hawkeye} to extract both static and dynamic CFGs from binaries (i.e. Windows, Linux and Android platforms). Embeddings of instructions are computed using word2vec whereas the final graph embedding is calculated using a custom GNN that leverages the word embeddings as nodes. Malware samples were collected from VirusShare and Androzoo, and benign examples for Windows and Linux platforms were collected from libraries. Using this cross-platform method, HawkEye reaches an accuracy of 96.82\% on Linux, 93.39\% on Windows, whereas 99.6\% accuracy is obtained on Android.

A similar CFG based on an assembly is employed in the paper \cite{wang2021hierarchical}. First, the semantic of functions is computed using random walk and the BERT \cite{devlin2018bert} language model. These embeddings are then assigned to the nodes of a FCG that represents a global view of the program. The importance between function nodes is then calculated with a GAT model, whose goal is to compute an attention score $e_{ij}$ for each connected pair of function nodes $i$ and $j$:
\begin{align*}
    e_{i j}=\operatorname{LeakyReLU}\left(\mathbf{a}^T\left[\mathbf{W} h_i, \mathbf{W} h_j\right]\right)
\end{align*}
where $[,]$ is the concatenation operation, $\mathbf{a}$ and $\textbf{W}$ respectively represent a trainable attention vector and a weight matrix. The features of nodes $i$ and $j$ are respectively represented here by $h_i$ and $h_j$. As this score is not normalized, a softmax activation is applied on all neighbors to obtain coefficients that are associated to probability distribution: 
\begin{align*}
    \alpha_{i j}=\operatorname{softmax}_j\left(e_{i j}\right)=\frac{\exp \left(e_{i j}\right)}{\sum_{k \in \mathcal{N}_i} \exp \left(e_{i k}\right)}
\end{align*}
where $\mathcal{N}_i$ is the neighborhood of node $i$. The updated representation $h_i^{\prime}$ of node $i$ can be obtained by gathering neighbor embeddings along with the calculated coefficients. The authors also leverage multi-head attention to calculate multiple representations that are then concatenated together:
\begin{align*}
h_i^{\prime}=\|_{k=1}^K \sigma\left(\sum_{j \in \mathcal{N}_i} \alpha_{i j}^k \textbf{W}^k h_j\right)+\mathbf{W}_{\mathbf{R}} h_i
\end{align*}
where $||$ represents the concatenation operation, $K$ is the number of attention heads and $\mathbf{W}_{\mathbf{R}} h_i$ is a trainable residual connection.
By leveraging the function-level and program-level embeddings with attention, the overall model achieved 90.88\% and 72.44\% F1-score on two private datasets.

In MalGraph \cite{ling2022malgraph}, both CFG and FCG are also leveraged together. Intra-procedural relations are captured with GraphSAGE from the CFG and the embeddings are attributed to nodes in a FCG whose embeddings are also computed with GraphSAGE to capture inter-procedural relations. Max-pooling transforms node embeddings into graph embeddings for downstream PE malware graph classification. All samples were collected from VirusShare and VirusTotal and IDA Pro was utilized for disassembling.

Although previous works are by definition detection methods, they provide poor insights on the actual patterns and areas in the graph that led to the final prediction.
CFGExplainer \cite{herath2022cfgexplainer} is an explainability framework specially designed to explain the predictions done by GNNs on malware classification tasks based on CFGs. This method identifies subgraphs in the CFG, that contribute to the final prediction of a given GNN. Concerning this particular task, CFGExplainer outperforms other explainability frameworks like GNNExplainer \cite{ying2019gnnexplainer}, SubgraphX \cite{yuan2021explainability} and PGExplainer \cite{luo2020parameterized}.

\subsubsection{FCG Approaches for Windows Malware Detection}
DLGraph \cite{jiang2018dlgraph} leverages static analysis to extract API calls and FCG from binaries. The model relies on a FCG that represents the interactions between functions from disassembled PE files, along with a vector of extracted Windows API calls. The node embeddings of the FCG are calculated using node2vec and are fed into a stacked denoising auto encoder (SDA) \cite{vincent2010stacked} to create a graph embedding vector. Similarly, a SDA takes as input the API vector and the two resulting vectors are then concatenated and passed into a softmax regression layer for classification, where an accuracy greater than 99\% is achieved.

In the paper \cite{oliveira2019behavioral}, a behavioral graph is constructed from API call sequences monitored during the execution of PEs in a sandbox environment. A DGCNN is then used to compute embeddings for graph classification. Based on their experiments, the authors released a public dataset made of 42,797 malware and 1,079 benignware API call sequences \cite{tqqm-aq14-19}. On the original imbalanced dataset, the proposed model achieves an F1-score of more than 99\%. 

SDGNet \cite{zhang2020spectral} similarly captures API calls along with attributes using dynamic analysis. Weighted graph normalization methods are utilized to transform the adjacency matrix into three symmetrical matrices that describe interactions of node information. A GCN-based model computes node embeddings for these matrices and all representations are merged into a final graph embedding that is leveraged for classification. A total of 8,909 labeled samples were collected from the Alibaba Cloud Malware Detection Base on Behavior dataset \cite{alibaba} and the final model achieves 97.3\% accuracy.

The reference \cite{li2022intelligent} uses a GCN on a directed cyclic graph that was pre-processed with Markov chain. The nodes represent API calls, whereas the edges $(u,v)$ are weighted according to the number of calls from $u$ to $v$. Malware samples used for evaluation are also collected dynamically using a sandbox environment in order to create a private dataset, on which a maximum accuracy of 98.32\% is reached.

DMalNet \cite{li2022dmalnet} also leverages dynamic analysis to build FCGs from API calls captured during the execution of PE binaries. Here, both API names and API arguments are considered. The embeddings of these attributes are learned with a custom GIN model, whereas more complex structural interactions between APIs are learned with attention using a GAT. After computing embeddings for both semantics, important information is captured with a gPool layer \cite{cangea2018towards,gao2019graph} for feature selection. More precisely, this pooling operation attributes to each node a projection score and selects top-$k$ nodes based on these scores. The gPool output of the GIN model is taken as input by the GAT to capture the interactions between API calls. A final accuracy of 98.43\% is obtained by leveraging a MLP for classification on a private dataset.

\subsubsection{Entity Graph Approaches for Windows Malware Detection}
The dynamic nature of communications between system entities provide valuable information to detect malicious behaviors. In the paper \cite{hung2019malware}, authors monitor such interactions using dynamic analysis. Directed multi-edge graphs are built from interactions between four system entities: processes, files, registry keys and network sockets. Edges represent data transmission between entities such as system calls. Concretely, a node is represented by a categorical value between 0 and 3, and an edge stores a feature vector containing the size of the transmitted data and the time when the action occurred. Representation learning is done using a GCN and an attention-based pooling function is used to transform node embeddings into fixed-size graph embedding vector that is classified by a feed-forward network. Experiments have been conducted on a private dataset with samples from VirusShare and the proposed solution achieved 86.22\% accuracy.

In MatchGNet \cite{wang2019heterogeneous}, malware detection is considered as the detection of a malicious process that behaves differently from benign processes. The authors first designed an invariant graph modeling technique to capture interactions in a heterogeneous graph that represents relations among system entities such as processes, files or sockets. A GNN-based encoder with attention learns the representations and a Siamese Network \cite{bromley1993signature} learns the similarity between known benign programs and new incoming programs. During inference, the similarity distance between these two programs results in a score that is utilized for final classification. This model can thus be trained using only benign examples. The final evaluation is performed on a real enterprise dataset composed of 300 million events recorded on Windows and Linux hosts.

The paper \cite{fan2020novel} represents malware behaviors as a weighted heterogeneous graph, where nodes are either an executable file (PE), a file, a file suffix or a module, and edges represent different (weighted) actions between entities. A custom model based on GraphSAGE and meta-paths is implemented to deal with the heterogeneity of the graph. This model achieved 91.56\% accuracy on a private dataset.

FewM-HGCL \cite{liu2022fewm} introduces a self-supervised method based on contrastive learning for few-shot malware variants detection. The authors construct a heterogeneous graph with 5 types of entities: process, API, file, signature, and network. API nodes are not only identified by their name but are also characterized by their category. API attributes are irregular by default and feature hashing \cite{weinberger2009feature} is used to transform them into compact fixed-length vectors. The idea behind contrastive learning is to use the natural co-occurrence associations in data as a substitute for ground truth labeled information. To perform contrastive learning, negative samples and positive samples are generated using different data augmentation techniques. Three distinct GAT models are then trained to respectively learn graph embeddings on the original graph, the positive graph and the negative graph. A discriminator aims to capture the similarity between the original graph and the positive graph along with the dissimilarity between the original graph and the negative graph. All self-trained embeddings are finally merged in a readout layer for downstream graph classification with an accuracy ranging from 85.73\% to 98.65\% on multiple datasets for malware variants detection.

\subsection{Windows Malware Datasets}
On Windows platform, a majority of works rely on private datasets constructed from public malware samples downloaded from VirusShare and VirusTotal. Using these data, the comparison between papers is ineffective. Other works evaluate their experiments on the Microsoft Malware Classification Challenge, which makes the performance comparison between papers possible. Datasets used in previous studies are reviewed in Table \ref{table:windowsdatasets}.

\begin{table*}[ht]
\tiny
\caption{Datasets employed in Windows malware detection studies.}
\centering
\begin{tabular}{@{}p{0.15\textwidth}p{0.3\textwidth}@{}p{0.15\textwidth}}\toprule
\textbf{Paper} & \textbf{Datasets} & \textbf{Performance} \\

\midrule

MAGIC \cite{yan2019classifying} &	MMCC & 99.25\% acc \\
HawkEye \cite{xu2021hawkeye} & VirusShare+AndroZoo & 96.82\%, 93.39\%, 99.6\% acc \\
Wang et al. \cite{wang2021hierarchical} &	VirusShare+VirusTotal &	90.88\%, 72.44\% F1 \\
MalGraph \cite{ling2022malgraph} & VirusShare+VirusTotal & 99.97\% acc \\
DLGraph \cite{jiang2018dlgraph} &	MMCC+VirusShare+KafanBBS &	>99\% acc \\
Oliveira et al. \cite{oliveira2019behavioral} &	VirusShare & \textasciitilde 99.4\% F1 \\
SDGNet \cite{zhang2020spectral} & Alibaba Dataset & 97.3\% acc \\
Li et al. \cite{li2022intelligent} &	VirusShare+VirusTotal &	98.32\% acc \\
DMalNet \cite{li2022dmalnet} &	VirusShare+VirusTotal &	98.43\% acc \\
MeQDFG \cite{hung2019malware} &	VirusShare &	86.22\% acc \\
MatchGNet \cite{wang2019heterogeneous} & Private & 96.53\% acc \\
MalSage \cite{fan2020novel} &	VirusTotal &	91.56\% acc \\
FewM-HGCL \cite{liu2022fewm} & \parbox{0.22\textwidth}{VirusShare,ACT-KingKong,Ember,\\API Call Sequences,BIG 2015}	 & 85.73\%-98.65\% acc \\

\bottomrule
\end{tabular}
\begin{tablenotes}
  \setlength\labelsep{0pt}
    \footnotesize
  
\end{tablenotes}
\label{table:windowsdatasets}
\end{table*}

\begin{microsoft*}
This dataset contains more than 20,000 malware samples that fall into nine families, namely Ramnit, Lollipop, Kelihos ver3, Vundo, Simda, Tracur, Kelihos ver1, Obfuscator.ACY and Gatak. For each binary, the dataset provides two data representations: the bytecode and the disassembly code (disassembled with IDA Pro). The assembly code can then be used to build attributed CFGs \cite{yan2019classifying} or FCGs \cite{jiang2018dlgraph}.
\end{microsoft*}

\begin{virushsharevirushtotal*}
It is common to download PE malware and Android malware from these two websites. Some works rely on dynamic analysis to run the downloaded malware into a cuckoo sandbox \cite{li2022intelligent,li2022dmalnet,hung2019malware}, whereas others build static CFGs and FCGs \cite{wang2021hierarchical,jiang2018dlgraph}.
\end{virushsharevirushtotal*}

\section{Graph-based Web Malware Detection} \label{sec:web}
Compiled binaries are not the only way to hide malware payloads with the intent to execute malicious activity. Malware are also present in web technologies and efficient cyberdefense systems should also be designed. Graph representation learning remains little used in the web-based malware detection, but we think that web technologies are by definition graph-oriented and they could be potential candidates to these graph learning techniques. In this section, we will cover several graph structures for web malware detection along with works involving graph representation learning.

\subsection{Web-based Graph Structures}
\begin{figure}[h]
\centerline{\includegraphics[width=0.8\textwidth,keepaspectratio]{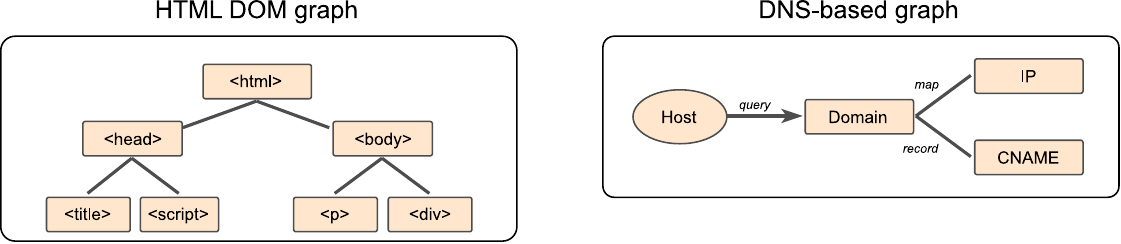}}
\caption{Example of Web-based graph structures built from HTML code (left) and DNS data (right).}
\label{fig:webmalware}
\end{figure}

The web contains a wide range of interconnected web pages accessible through Internet. All these pages along with the fundamental components that compose the structure of Internet are likely to hide malware used by attackers to fool unaware web visitors. Detecting such malicious behaviors can be achieved by direclty analyzing the source code of pages or the communications between services present on the web. In this survey, we focus our study on techniques employed to detect malware activities from source code, DNS communications and network flows. For the representation of web malware as code, web pages' content is generally fetched in order to build a hierarchical graph from HTML or JavaScript code. The DNS scene can also be modeled as a graph to excavate useful associations among domains, as shown in Fig. \ref{fig:webmalware}. In this case, the graph is generally heterogeneous as it models complex interactions between different types of entities such as hosts, IP addresses, segments, etc. A graph representation learning model can then learn malicious code patterns hidden in the code of untrusted websites, or detect anomalous DNS communications likely to be attacks such as DNS spoofing or DNS flood attacks. As for Android and Windows malware detection, network flows may also be used to model IP communications between clients and to detect network-level attacks like botnet and DDoS.

\subsection{Web-based Approaches}
Recent approaches for web-based malware detection with graph representation learning are presented below and summarized in Table \ref{table:websummary}.

\begin{table*}[ht]
\tiny
\caption{Summary of Web-based malware detection approaches leveraging graph representation learning.}
\centering
\begin{tabular}{@{}p{0.07\textwidth}p{0.05\textwidth}p{0.11\textwidth}p{0.09\textwidth}p{0.12\textwidth}p{0.19\textwidth}p{0.04\textwidth}p{0.15\textwidth}@{}}\toprule
\textbf{Data} & \textbf{Analysis} & \textbf{Graph type} & \textbf{Classification} & \textbf{Learning} & \textbf{Models} & \textbf{Year} & \textbf{Paper} \\

\midrule

\multirow{4}{*}{\textbf{Code}} &  Static & Attributed, Heterogeneous & Graph & Supervised & TAGCN, RNN & 2021 & Ouyang et al. \cite{ouyang2021phishing} \\

& Hybrid & Attributed & Graph & Semi-supervised & GCN, Random Forest & 2022 & PhishGNN \cite{secrypt22} \\
& Static & Attributed & Node & Semi-supervised & GCN, Residual connections, word2vec & 2022 & GraphXSS \cite{liu2022graphxss} \\

 & Static & Attributed & Graph & Supervised & Gated GNN, GAT, word2vec & 2022 & JStrong \cite{fang2022jstrong} \\

\cmidrule{1-8}
\multirow{3}{*}{\textbf{DNS data}} & \multirow{3}{*}{Dynamic} & Attributed & Node & Supervised & DeepWalk & 2019 & He et al. \cite{he2019malicious} \\

&  & Heterogeneous & Node & Supervised & Heterogeneous GCN, Meta-path, Random Walk & 2020 & Deepdom \cite{sun2020deepdom} \\

&  & Attributed, Heterogeneous & Node & Semi-supervised & Custom GNN with attention & 2021 & GAMD \cite{zhang2021attributed} \\

\cmidrule{1-8}
\multirow{4}{*}{\textbf{Flows}} & \multirow{4}{*}{Dynamic} & Graph & Node & Supervised & GCN & 2020 & Zhou et al. \cite{zhou2020automating} \\

&  & Graph & Node & Supervised & Inferential SIR\_GN & 2022 & isirgn1 \cite{carpenter2021detecting} \\

&  & Heterogeneous & Node & Semi-supervised & GCN, Meta-path & 2020 & Bot-AHGCN \cite{zhao2020multi} \\

&  & Attributed & Graph & Supervised & GIN & 2022 & GraphDDoS \cite{li2022graphddos} \\

\bottomrule
\end{tabular}
\begin{tablenotes}
  \setlength\labelsep{0pt}
    \footnotesize
\end{tablenotes}
\label{table:websummary}
\end{table*}

\subsubsection{Code-based Approaches for Web Malware Detection}
The inner structure of web pages contains important indicators that may be represented as a graph for downstream graph ML tasks. Ouyang et al. \cite{ouyang2021phishing} suggests to fetch the HTML content of phishing web pages to build a heterogeneous graph on which learn the malicious structures that may help the detection of phishing attacks. Here, the HTML files are parsed into DOM trees, where a node represents a HTML tag \cite{tags} and edges are links between the tag and its inner tags. Leaf nodes are especially considered, as they store the actual string content that is located and displayed on the web page. A RNN is used at node-level to encode type-specific features based on the attributes and the text content. To capture long-range relations, the authors used the Topology Adaptive Graph Convolutional Network (TAGCN) \cite{du2017topology} to aggregate neighbors from different hops instead of the direct neighborhood used in traditional GCNs. All embeddings are then reduced with max-pooling and fed into a fully-connected layer for graph classification. The model is trained in an end-to-end manner with cross-entropy loss on a private dataset created from public phishing repositories, with a 95.5\% accuracy.

On the other hand, PhishGNN \cite{secrypt22} leverages the hyperlink structure of web pages, extracted from HTML content. Here, the graph is built by crawling the outgoing links of a HTML page, up to a certain depth. Each node represents a URL that is mapped to a vector of 25 hand-designed features automatically extracted during crawling from the URL text, the DOM content and the domain. The proposed framework first trains a random forest algorithm with the 25 features of all known websites, and then predicts the label of the outgoing links extracted with the crawler. A message-passing GNN such as GCN is then used on top of the graph containing the random forest predictions, and graph embeddings are aggregated with graph pooling for downstream classification. The combination of random forest predictions with a GNN achieves a clear classification improvement when compared to the performance of each algorithm separately. In the same way as \cite{ouyang2021phishing}, the performance of the model is evaluated on a private dataset, where PhishGNN reaches an accuracy of more than 99\%. 

Against malicious XSS payloads, authors in \cite{liu2022graphxss} propose to first preprocess payload samples with word2vec and further leverage the relations between words by creating a graph on which a GCN learns embeddings. A residual connection is employed in the GCN propagation function to accelerate the convergence of the model. This implementation is then evaluated on a private XSS payload dataset, with a 99.6\% prediction accuracy.

JStrong \cite{fang2022jstrong} leverages graph representation learning to detect malicious JavaScript (JS) code. The authors study the performance of multiple graph structures to effectively represent the semantic of the JS code. Notably, word2vec along with a Gated GNN and a GAT model are leveraged to compute the embeddings of different graphs, namely an Abstract Syntax Tree (AST), a CFG, a PDG and an Object Dependence Graph (ODG) \cite{li2022mining}. Embeddings are then aggregated with mean pooling and classified in graph classification setting, where best performance is achieved by applying the final architecture on a pruned PDG, which retains the most important information. Indeed, the authors explain that the PDG extracts more semantic information than an AST or a CFG due its ability to model data flow information and statement execution order.

\subsubsection{Domain-based Approaches for Web Malware Detection}
For the detection of malicious domains, the paper \cite{he2019malicious} proposes a lightweight approach that considers domains from passive DNS data. First, a domain relationship graph is built from domains and A records, where each node represents a domain and an edge exists if two domains resolve to the same IP address. Node embeddings are then computed using DeepWalk and the prediction is enhanced with hand-designed DNS features.

Deepdom \cite{sun2020deepdom} applies a custom GCN on meta-paths extracted from a HIN containing entities such as clients, domains, IP addresses, and multiple relations like query, register and record. The proposed method has the ability to support inductive node embedding, and can thus generalize to unseen DNS nodes.

Another approach to detect unwanted domains is suggested in the paper \cite{zhang2021attributed}. Here, the DNS interactions extracted from a private university network are also modeled with a HIN, where nodes are either a host, a domain or a resolved-IP, and edges could be a request or a resolution relation. Self-attention mechanism is applied directly on the heterogeneous edges, in contrast to HAN \cite{wang2019heterogeneous} that uses this mechanism on meta-path scheme. With this technique, the representation of different types of neighbor nodes is projected in the embedding space of the specific target node type. Node embedding of malicious domains can then be detected in embedding space using a downstream fully-connected layer.

\subsubsection{Network Flow Approaches for Web Malware Detection}
Network monitoring tools are effective to capture malicious traffic that may come from malware programs. 
In the paper \cite{zhou2020automating}, authors propose to automate the detection of botnets by leveraging network traces between hosts with a GNN-based method.
The task is here to classify malicious nodes, namely an IP address that participates in the botnet attack. A GCN model first computes the node embeddings across all the graph and malicious botnet nodes are then classified using a neural network at node-level. In botnets, infected bots receive commands from either centralized command-and-control (C\&C) or decentralized peer-to-peer architectures (P2P). This technique respectively aims to detect C\&C and P2P botnets with 99.03\% and 99.51\% accuracy. For this evaluation, a private dataset was created based on botnet topologies and background network traffic.

Another approach trained on background traffic with embedded synthetic botnet topologies is considered in reference \cite{carpenter2021detecting}. Here, the Inferential SIR\_GN model is used to generalize on unseen and very large graphs. Indeed, network-based graphs can rapidly grow in size and adapted models are required to deal with these high-dimension structures. The node embeddings computed by SIR\_GN are then fed into a standard neural network to classify botnet nodes.

Zhao et al. \cite{zhao2020multi} represents flows as a heterogeneous graph where nodes are flow entities and edges are events between flows.
The nodes are also attributed with features such as timestamp and user-agent. Meta-paths are hand-designed to extract semantic from the heterogeneous graph. Then, a weighted similarity graph is built by computing similarity between node pairs and a GCN computes the embeddings. As in previous approaches, botnet nodes are detected using a neural network at node-level.

Network flows also provide useful information to detect DDoS attacks. In the paper \cite{li2022graphddos}, GraphDDoS aims to detect low-rate and high-rate DDoS attacks by considering the relationship between flows along with the relationship of packets from a single flow. First of all, an endpoint graph is constructed by dividing packets into two groups based on the source and destination IP addresses. The GIN model then performs message-passing between every nodes and computes embeddings. The task here is to classify DDoS attack graphs so the node embeddings are passed into a readout layer to perform graph classification.

\subsection{Web Malware Datasets}
In this section, we present web-based datasets on which graph structures can be built for graph representation learning.

\begin{table*}[ht]
\tiny
\caption{Datasets employed in Web malware detection studies.}
\centering
\begin{tabular}{@{}p{0.15\textwidth}p{0.3\textwidth}@{}p{0.17\textwidth}}\toprule
\textbf{Paper} & \textbf{Datasets} & \textbf{Performance} \\

\midrule

Ouyang et al. \cite{ouyang2021phishing} &	OpenPhish+PhishTank+TrancoTop1M &	95.5\% acc \\
PhishGNN \cite{secrypt22} &	OpenPhish+PhishTank+AlexaTop1M &	99.7\% acc \\

GraphXSS \cite{liu2022graphxss}	& XSSed & 99.6\% acc \\
JStrong \cite{fang2022jstrong} & Petrak+GeeksOnSecurity+VirusTotal & 99.95\% acc \\
He et al. \cite{he2019malicious} & Various datasets+AlexaTop1M & 94\% acc \\
Deepdom \cite{sun2020deepdom} & Private & 97.91\% acc \\
GAMD \cite{zhang2021attributed}	& Private & 92.77\% acc \\

Zhou et al. \cite{zhou2020automating} &	Private & 99.03\%, 99.51\% acc \\
isirgn1 \cite{carpenter2021detecting} &	CAIDA+Synthetic samples	& 97.85\%-99.78\% F1 \\
Bot-AHGCN \cite{zhao2020multi} &	CTU-13, Private & 98.27\%, 98.22\% micro-F1 \\
GraphDDoS \cite{li2022graphddos} &	CIC-IDS-2017, CIC-DoS-2017 &99.59\%, 94.56\% F1 \\

\bottomrule
\end{tabular}
\begin{tablenotes}
  \setlength\labelsep{0pt}
    \footnotesize
\end{tablenotes}
\label{table:webdatasets}
\end{table*}

\begin{phishtank*}
These two websites provide an updated list of known malicious URLs that is frequently updated by the community. Phishing URL detection is then possible either by directly extracting features from the raw URL as text, or by crawling the webpage's content if the domain still exists.
\end{phishtank*}

\begin{alexa*}
Provide benign URLs from the top 1 million sites on Internet ranked by traffic. Often used in combination with malicious URLs from PhishTank and OpenPhish.
\end{alexa*}

\begin{xssed*}
An online webpage providing an updated list of XSS payloads and XSS vulnerable websites. This page was created in early 2007 with the scope of increasing security and privacy on the web, and remains today the largest online archive of XSS vulnerable websites \cite{xssed}. 
\end{xssed*}

\begin{petrak*}
Two datasets hosted on GitHub, containing malicious JS file samples. Petrak's dataset contains almost 40,000 JS malware samples, whereas the second contains malware samples divided into 1,156 HTML files, 1,357 JS files and 33 skipped files.
\end{petrak*}

\begin{caida*}
Between 2008 and 2019, the Center for Applied Internet Data Analysis (CAIDA) captured passive network traces from high-speed monitors on a business backbone link. Pcap files allow access to hundreds of Gigabytes of requests that were logged over the course of these years. All these traces provide a good solution to model background traffic in synthetic network datasets \cite{zhou2020automating}. 
\end{caida*}

\begin{ctu*}
This dataset was made public by CTU University, Czech Republic. It contains network traffic captures from benign activity and from 13 botnet attack scenarios. Packets are available in pcap format and flows are in Netflow format and captured with Argus. In total, more than 850M packets and around 20M bi-directional flows are proposed in the dataset.
\end{ctu*}

\begin{cic2017*}
CIC-IDS2017 is a network dataset suggested by the Canadian Institute of Cybersecurity (CIC). It consists of benign and offensive network flows. Each flow is associated with 80 features collected over the course of 5 days in a controlled setting using CICFlowMeter. These data are available in pcap and CSV format. Seven types of web attacks are represented, namely Brute Force, HeartBleed, Botnet, DoS, DDoS, Web Attack, and Infiltration.
\end{cic2017*}

\begin{cicdos*}
This dataset provides network traces from common application layer DoS attacks simulated in a testbed environment. The victim host is a webserver running Apache Linux and the attacker is supposed to be non-oblivious, meaning that he knows to optimize traffic to maximize the attack damage. The resulting experiment lasts for 24 hours and the final dataset results in 4.6GB of data.
\end{cicdos*}

\section{Adversarial Attacks} \label{adversarial}
Despite the capabilities of machine learning in classification tasks, these techniques are not immune to adversarial attacks, that aim to disturb the predictions of the model by introducing adversarial examples, crafted from small perturbations in the input. We review in this section background knowledge on adversarial attacks along with existing approaches against traditional and graph-based malware detection.

\subsection{Background}
Adversarial attacks have seen great success notably in the computer vision domain \cite{akhtar2018threat}, where the goal is to craft adversarial images that the model will misclassify by predicting a wrong label. For a given classification model $f$ denoted as $f: x \rightarrow y$ that predicts a label $y \in \mathbb{Y}$ given the features $x \in \mathbb{X}$ of an input example $z \in \mathbb{Z}$, we denote two categories of adversarial attacks \cite{ling2023adversarial,demetrio2021adversarial}. A feature-space attack aims to craft adversarial features $x^{\prime} \in \mathbb{X}$ (e.g. a modified FCG or CFG) such that the distance between $x$ and $x^{\prime}$ in feature-space is minimized, and such that the model $f$ predicts a label $y^{\prime} \in \mathbb{Y}$ different from $y$. However, a problem-space attack works directly on the real-world input $z$ instead of the features $x$. The goal then becomes to minimize the cost between $z$ and an adversarial example $z^{\prime}$ (e.g. modified source code), such that $f$ predicts another label $y^{\prime}$. We can further classify adversarial attacks by the prior knowledge acquired by the attacker. White-box attacks assume that the attacker has full knowledge of the target model $f$, namely he knows about the architecture, the parameters, etc. In contrast, black-box attacks refer to scenarios where only the output prediction is known by the attacker, making these attacks more difficult to succeed  but also more likely to be faced in real-world applications. Other methods called gray-box attacks, live at the intersection between black- and white-box methods, where the attacker has knowledge of some prior knowledge that shall be defined depending on the use case.

\subsection{Adversarial Attacks and Malware Detection} \label{sec:adversarial_malware}
In the case of malware, adversarial attacks aim to craft new malware examples that preserve maliciousness while 
misleading the classification of the model. Formally, given an input malware $z$ such as a PE or an APK, we want to find either a modified version of its compiled code $z^{\prime}$ or a modified version of its graph representation $x^{\prime}$ that will not be detected by the model. However, these requirements imply multiple constraints that are hard to be fulfilled in the case of malware adversarial attacks. Indeed, as explained by Ling et al. \cite{ling2023adversarial}, adversarial attacks have been successfully applied to image classification as it is easy to retrieve a corresponding image $z^{\prime}$ from an adversarial feature $x^{\prime}$ because an image can be simply represented as a 2D-array of pixels. In other words, a differentiable and bi-injective inverse feature mapping function $\phi^{-1}$ can be approximated to map features from the feature space to an image in problem space. However, retrieving the original malware code $z^{\prime}$ from a feature representation $x^{\prime}$ (i.e. finding a similar inverse function) is much more challenging as the reconstructed input needs to fulfill multiple conditions to remain executable \cite{pierazzi2020intriguing}. Notably, the generated adversarial example needs to respect a specific format such as PE or APK, but also needs to preserve the malicious payload while still being executable without error. Furthermore, in a black-box scenario, the attacker does not know beforehand the feature representation taken as input by the detection model, which further complicates the adversarial process.

Despite these complicated requirements, researchers found adversarial attacks that can be employed to detect malware. To evade raw bytes-based malware detection models, works \cite{kreuk2018deceiving} and \cite{suciu2019exploring} append an adversarial sequence of bytes to the malware. Other works prefer to modify regions in the PE header \cite{demetrio2019explaining} or extend the DOS header \cite{demetrio2021adversarial}. However, these techniques are ineffective for higher-level representations such as those based on API calls. For this purpose, many works insert additional API calls in feature space to add noise in the representation and evade the detection systems \cite{chen2017adversarial,hu2018black,rosenberg2018generic,kawai2019improved,fadadu2020evading,rosenberg2020query,hu2023generating}. In other works, reinforcement learning (RL) is leveraged to manipulate the original malware in order to evade detection while maintaining a correct format and semantic \cite{anderson2017evading,wu2018enhancing,fang2019evading}.

\subsection{Adversarial Attacks on Graph-based Malware Detection} \label{sec:adversarial_graph_malware}
Adversarial attacks are inherently dependent on the data representation taken as input by the model. When working with GNNs, attackers thus need to consider the graph representation of the data, leading to adversarial attacks specifically designed for graph-based detection systems. Literature presents numerous papers that apply such attacks to GNN classifiers by either modifying node and edge features, or by directly manipulating the graph structure with actions such as removing or adding nodes and edges \cite{sun2018adversarial,zugner2018adversarial,dai2018adversarial}.
In the case of malware, removing nodes or edges from graph structures such as FCG or CFG is not appropriate as it would not preserve the functionality of the program. The adversarial attack should also be efficient on graph classification tasks, as a large majority of works leverage this task for malware detection.

Two such adversarial approaches specifically designed against call graph-based malware detection are proposed by Xu et al. in MANIS \cite{xu2020manis}. The first method aims to pick the $n$-strongest nodes from the graph, which are the nodes that have the most influence over their neighbor nodes. They are then inserted in the input graph until evasion has succeeded. The second method relies on the direction of the gradient to guide the insertion of new nodes. The advantage of these methods is that they produce a valid binary that preserves the given format (e.g. PE or APK). On the Drebin dataset, 72.2\% misclassification rate is achieved with the $n$-strongest nodes method, whereas the gradient-based proposition reaches 33.4\% misclassification rate under the white-box setting. Similar results are also obtained in gray-box setting.

In the paper \cite{zhao2021structural}, authors propose a structural attack for APK-based FCGs that aims to address the inverse mapping problem \cite{pierazzi2020intriguing}, that consists in retrieving a valid malware in problem-space from the modified malware in feature-space (see Section \ref{sec:adversarial_malware}). The proposed method works in white-box setting, and leverages reinforcement learning along with heuristic optimization to perform graph modifications such as inserting and deleting nodes, or adding edges and rewiring. The performance of this solution has been evaluated on 30k APKs from Androzoo with over 90\% attack success rate in feature space and up to 100\% attack success rate in problem space.

Another adversarial method based on reinforcement learning is introduced in reference \cite{zhang2022semantics} to evade GNN-based malware detection from CFGs. A deep RL agent is trained to insert semantic NOPs (no-operations) in CFG  basic blocks extracted from PE malware. This technique has the faculty to preserve the semantic and format of the original file, while evading GNN classifiers in black-box setting with nearly 100\% attack success rate on CFGs constructed with Angr \cite{angr} from samples collected on VirusShare and from the VXHeavens dataset \cite{tan2016artificial}.

An adversarial attack for GNN-based APK malware detection has been introduced in the work \cite{yumlembam2022iot} to measure the robustness of the proposed detection model. The attack is based on a VGAE, that aims to effectively add nodes and edges to a FCG in order to fool the GNN classifier, in a black-box setting. This adversarial approach has been applied to the original GNN model to further improve the robustness of the detection system.



\section{Challenges and Directions} \label{sec:future}
Graph representation learning has only recently been applied to malware detection. Therefore, there are still many challenges to achieve resilient malware detection methods. Consequently, we provide some future directions that could improve research in this area: 

\begin{itemize}
    \item While many papers reach good performance on malware detection using graph representation learning techniques, these models are usually evaluated on distinct examples. Indeed, some popular datasets exist, but they are often supplemented by additional samples extracted from public repositories such as Google Play Store, VirusShare and VirusTotal. These new samples make the comparison between papers inefficient as training and testing steps are not performed on the same malware examples, thus leading to different performance evaluations. We think that a large and diversified baseline dataset would be needed for future work, with the aim to effectively compare the metrics of different models.
    \item The robustness of current approaches based on GNNs is uncertain. Most current works rely solely on the code extracted from APKs using static analysis. However, detecting obfuscated malware by only using its code is a challenging task \cite{balakrishnan2005code}. Additional efforts using hybrid approaches on graphs could improve the robustness of these techniques, but this remains a scarcely explored direction. \\Furthermore, attackers may try to bypass the detection capabilities of the model by leveraging adversarial attacks. However, defenses against these attacks remains little studied in the field of GNNs and even less when applied to graph-based malware detection. Existing adversarial approaches presented in Section \ref{sec:adversarial_graph_malware} have proven remarkable results in fooling the predictions of GNN-based classifiers even in black-box scenarios, meaning that important efforts are still necessary to obtain robust methods.
    \item One of the drawbacks of using deep models is that they are not amenable to interpretability since they function as black boxes. However, understanding the reasons of a predictive model is of main importance, especially in the field of cybersecurity, where analysts should be able to understand the security-related decisions taken by algorithms. Explainability techniques currently exist to provide insights on the predictions performed by deep architectures such as GNNs \cite{ying2019gnnexplainer}. However, very few works leverage these techniques to further improve the explainability of malware predictions with GNNs \cite{warmsley2022survey} and further research in this direction could be very useful to the fields of malware detection and analysis.
\end{itemize}
Furthermore, widely used GNN architectures may not be optimized for the particular task of malware detection, as these models were not specifically designed for this purpose. This means that significant research work could be undertaken to discover new GNN models dedicated to the representation of malware.

\section{Conclusion} \label{conclusion}
In this paper, we provide an in-depth review of graph representation learning techniques applied to the detection of Android, Windows and Web malware. We first introduced fundamental knowledge to understand graph-based learning methods along with the graph structures commonly employed in malware detection. We reviewed and classified state-of-the-art works in a comprehensive way and provide descriptions and insights on the datasets that can be leveraged to represent malware as graphs. We notably found that most existing techniques can be represented under a same architecture based on graph classification, which is presented and used as reference in this survey.
We also discovered that many recent works prefer leveraging GNNs in combination with word embedding techniques to learn the semantic of disassembled code along with the structural patterns of existing malware. This survey also shows that effective adversarial attacks can be used by attackers in an attempt to fool graph-based detection systems. The analysis of recent papers demonstrates the promising future of graph ML methods applied to malware detection, and as a result, we have provided future research directions based on the current challenges that can be addressed.


\bibliographystyle{unsrt}
\bibliography{biblio}

\end{document}